\documentclass[prd,showpacs,preprintnumbers,twocolumn,superscriptaddress,floatfix]{revtex4}
\usepackage[utf8]{inputenc}
\usepackage{bm,amsmath,amssymb}
\usepackage{graphicx}
\usepackage{subfigure}
\usepackage{color}
\usepackage{hyperref}
\usepackage{multirow}
\usepackage{lineno}
\usepackage{soul}

\let\oldalign\align
\def\align{\linenomath\oldalign}

\newcommand{\Dov}{D_\text{ov}}

\begin{document}

\title{Microscopic Origin of \boldmath{$U_A(1)$} Symmetry Violation in the High Temperature Phase of QCD}

\author{Viktor Dick}
\affiliation{Fakult\"at f\"ur Physik, Universit\"at Bielefeld, D-33615 Bielefeld,
Germany}
\author{Frithjof Karsch}
\affiliation{Fakult\"at f\"ur Physik, Universit\"at Bielefeld, D-33615 Bielefeld,
Germany}
\affiliation{Brookhaven National Laboratory, Upton, NY 11973, USA}
\author{Edwin Laermann}
\affiliation{Fakult\"at f\"ur Physik, Universit\"at Bielefeld, D-33615 Bielefeld,
Germany}
\author{Swagato Mukherjee}
\affiliation{Brookhaven National Laboratory, Upton, NY 11973, USA}
\author{Sayantan Sharma}
\affiliation{Brookhaven National Laboratory, Upton, NY 11973, USA}

\begin{abstract}

We investigate the low-lying eigenmodes of the Dirac matrix with the aim to gain more insight 
into the temperature dependence of the anomalous $U_A(1)$ symmetry. We use 
the overlap operator to probe dynamical QCD configurations generated with 
(2+1)-flavors of highly
improved staggered quarks. We find no evidence of a gap opening up in the infrared
region of the eigenvalue spectrum even at $1.5\,T_c$, $T_c$ being the chiral crossover
temperature. Instead, we observe an accumulation of near-zero eigenmodes. We
argue that these near-zero eigenmodes are primarily responsible for the anomalous
breaking of the axial symmetry still being effective. At $1.5\,T_c$, these near-zero 
eigenmodes remain
localized and their distribution is consistent with the dilute instanton gas
picture. At this temperature, the average size of the instantons is $0.223(8)\,\text{fm}$ and
their density is $0.147(7)\,\text{fm}^{-4}$. 
    
\end{abstract}


\maketitle
\section{Introduction} 
\label{sc:intro}

Owing to the near-degeneracy and smallness of the up and down quark masses, the
Quantum Chromodynamics (QCD) Lagrangian possesses an approximate $U_L(2)\times U_R(2)
\equiv SU_L(2)\times SU_R(2)\times U_V(1)\times U_A(1)$ symmetry.  The fact that we
do not see parity doublet hadrons in our world implies that the  $SU_L(2)\times
SU_R(2)$ chiral symmetry is spontaneously broken down to the $SU_V(2)$ isospin symmetry of
the vacuum. It is well known from first principle lattice QCD studies that above the
chiral crossover~\cite{milceos,bnlbieos,bmweos,dwfTc} temperature of $T_c=154(9)$\,MeV 
\cite{hisqeos} the chiral symmetry of QCD gets restored. 

On the other hand, the axial $U_A(1)$ symmetry of the QCD Lagrangian is always broken due to
the presence of quantum fluctuations. This gives rise to the well-known anomalous
non-conservation of the axial current \cite{abj,fujikawa}. The explicit violation of
the global $U_A(1)$ symmetry is due to the presence of topologically nontrivial gauge field
configurations \cite{thooft1}. Although 
$U_A(1)$ is not an exact symmetry of QCD, the magnitude of its breaking near $T_c$ is
expected to influence the nature of the chiral phase transition in the limit of two
vanishingly small light quark masses.  Perturbative renormalization group studies of model
quantum field theories with the same global symmetries as QCD suggest that if
$U_A(1)$ is not effectively restored at $T_c$, the chiral phase transition is of
second order, belonging to the 3-dimensional $O(4)$ universality class
\cite{pw,bpv,pv,gr}. If the axial symmetry gets effectively restored for $T\sim T_c$,
the chiral phase transition can be either of  first order \cite{pw,bpv} or of second
order with the symmetry breaking pattern $U_L(2)\times U_R(2) \to U_V(2)$
\cite{pv,gr}.
In order to resolve the nature of the phase transition of QCD with two light quark flavors 
it is thus important to understand the significance of the anomalous $U_A(1)$ in the high 
temperature phase.

At low temperatures, $U_A(1)$ is also broken explicitly by the presence of a
non vanishing vacuum chiral condensate. In the chirally symmetric phase,
the vacuum condensate vanishes and the mechanism of global $U_A(1)$ breaking can be
studied directly. The microscopic mechanism for $U_A(1)$ breaking in the chirally
symmetric phase of QCD presents an intriguing puzzle. The chiral condensate, which is
the order parameter related to the restoration of chiral symmetry in QCD
with massless quarks, can be expressed in terms of the eigenvalues $\lambda$ of the
Dirac operator as
\begin{align}
    \label{eqn:ppbarc}
    \langle\bar\psi\psi\rangle\overset{V\rightarrow \infty}{\rightarrow}\int_0^\infty 
    \!\mathrm{d} \lambda \frac{2m~\rho(\lambda,m)}{\lambda^2+m^2} \;,
\end{align}
where $\rho(\lambda,m)$ is the eigenvalue density. On the
other hand, $U_A(1)$ is not a global symmetry, so one cannot define a corresponding
order parameter. For two light quark flavors, an approximate restoration of $U_A(1)$
would result in the degeneracy of the correlation functions of the pion and the scalar
iso-triplet delta meson \cite{shuryakw}. Specifically, the difference of the
integrated correlation functions of these mesons in terms of the eigenvalues of the 
Dirac operator is
\begin{align}
    \label{eqn:chipd}
    \chi_\pi-\chi_\delta &= \int\!\mathrm{d}^4 x~ \left[\langle i\pi^+(x)i\pi^-(0)\rangle
-\langle \delta^+(x)\delta^-(0)\rangle\right] \notag \\
&\overset{V\rightarrow \infty}{\rightarrow}
\int_0^\infty\! \mathrm{d} \lambda \frac{4m^2~\rho(\lambda,m)}{(\lambda^2+m^2)^2}~.
\end{align}
In the limit of vanishingly small quark mass $m$ and infinite volume $V$, the chiral
condensate is proportional to the density of near-zero eigenvalues in accordance with the
Banks-Casher relation~\cite{banks}, $\langle\bar\psi\psi\rangle=\pi\rho(0,0)$. For
$T\gtrsim T_c$, chiral symmetry gets restored and the chiral order parameter
$\langle\bar\psi\psi\rangle$ vanishes implying that $\rho(0,0)$ must also vanish. Motivated by the 
free theory limit at finite temperature, where the spectral density $\rho(\lambda,0)$ has a gap 
up to the lowest fermion Matsubara frequency,
$0\le\lambda<\pi T$, one possibility by which chiral symmetry restoration may occur
in the chiral limit is through the generation of a gap in the infrared part of the
eigenvalue spectrum. Such a scenario, however, would also lead to  the vanishing of 
$\chi_\pi-\chi_\delta$, i.e. to the effective restoration of both chiral and $U_A(1)$
symmetry.

In fact, more rigorous calculations based on chiral Ward identities for up to 4-point
correlation functions show \cite{aoki} that if the eigenvalue density for QCD with two light quark 
flavors is an analytic function in $m^2$, it must have the form $\lim_{m\to0} \rho(\lambda,m) \sim
\lambda^3 + \mathcal{O}(\lambda^4)$ in the chirally symmetric phase, 
similar to that for the free theory at $T=0$. It
was further shown \cite{aoki} that, in this case all correlation functions up to 6-point
which are related through $U_A(1)$ symmetry will be degenerate, making the anomalous
breaking of $U_A(1)$ invisible in these correlation functions. Thus, if
$U_A(1)$ breaking is finite through the nondegeneracy of the 2-point
correlation functions such as $\chi_\pi-\chi_\delta$, the eigenvalue density must be
nonanalytic in $m^2$.  Two such possible forms of the infrared eigenvalue spectrum,
compatible with $\langle\bar\psi\psi\rangle=0$ but $\chi_\pi-\chi_\delta\ne0$ for $m\to0$,
have been speculated in \cite{dw1}, namely
$\rho(\lambda,m)\sim m^2\delta(\lambda)$ and $\rho(\lambda,m)\sim |m|$. The functional form of
the infrared eigenvalue density in the chirally symmetric phase of QCD remains 
an open and interesting theoretical question.

The global $U_A(1)$ breaking at $T=0$ is intimately connected to the presence of topologically
nontrivial configurations of the QCD gauge fields~\cite{thooft1}. It is well known that 
localized topological structures like instantons give rise to zero modes
of the Dirac operator and the corresponding wavefunctions remain localized~\cite{thooft2}. 
The occurrence of near-zero modes can possibly also be traced back to the underlying topology of the gauge
field configurations. For example, the particular form of the eigenvalue density of the Dirac operator
$\rho(\lambda,m)\sim m^2\delta(\lambda)$ 
in the infrared can be motivated  from the fact that a small shift from zero of
the near-zero modes resulting from the weak interactions among widely separated
instantons and antiinstantons can be neglected, leading to a $\delta(\lambda)$ behavior. 
The $m^2$ factor naturally arises from the two light fermion
determinants. At high enough temperatures, it has been shown that the dilute gas of instantons 
is a reasonable description of pure gauge theory~\cite{gpy,ehn}. Within this approximation the instanton density is 
suppressed with decreasing value of the gauge coupling and eventually vanishes
when $T\to\infty$ \cite{gpy}. This dilute gas model is expected to be a good description 
of the high temperature phase of QCD as well~\cite{gpy}. The $U_A(1)$ breaking can be explained within such a model 
only for small enough values of the
gauge coupling, i.e. at sufficiently high temperature and for small sizes of
the instantons. It is however unclear whether such a mechanism can explain the $U_A(1)$
breaking for the more relevant temperature range from $T_c$ to a few times $T_c$. Near
$T_c$ the instantons and antiinstantons may not be widely separated and
weakly interacting, as described by the Instanton Liquid Model (ILM)
\cite{shuryak}. In this model, chiral symmetry breaking arises due to the fermion
modes associated with strongly interacting and overlapping instantons. As the
temperature is increased, it was proposed that there is a transition from a liquid
phase of disordered instantons and antiinstantons to a phase of instanton-antiinstanton
molecules \cite{shuryakilg1,sveb}. The chiral symmetry restoration at
finite temperature may not necessarily be due to the suppression of the instantons
\cite{shuryakilg2} but rather due to the temperature dependence of the fermion
determinant, which favors polarized instanton-antiinstanton molecules. Thus, it is
entirely possible that for $T\lesssim T_c\lesssim 2\,T_c$ other nonperturbative
mechanisms responsible for $U_A(1)$ breaking may generate an accumulation of
near-zero modes leading to a more complex form of the infrared eigenvalue spectrum. 

Since topological structures in QCD are inherently nonperturbative, lattice QCD
techniques are ideally suited to address issues related to the $U_A(1)$.
Anomalous $U_A(1)$ breaking at high temperature was studied on the lattice by looking
at the nondegeneracy of the 2 point correlation functions $\chi_\pi-\chi_\delta$
using staggered fermion formulation, which preserves a remnant of the continuum
chiral symmetry on the lattice, in \cite{shailesh} and more recently with an improved
staggered fermion formulation in \cite{cheng}. In both cases, it was observed that 
$U_A(1)$ was not effectively restored at $T\gtrsim T_c$. Recently, this issue was revisited in 
a study of the infrared eigenvalue spectrum of highly improved staggered quarks
(HISQ) \cite{hisq} and a similar conclusion was reached \cite{hiroshi}.  However, for
staggered fermions the connection between topology and fermion zero modes and hence
the reproduction of the index theorem is a very subtle issue \cite{adams}. Attempts
to study this problem with fermions with exact chiral symmetry on the lattice have
produced juxtaposing results. Studies with domain wall fermions \cite{dw1,dw2}
but with a heavier than physical pion mass of $200$ MeV support the scenario that $U_A(1)$ remains 
broken for $T\gtrsim T_c$. The eigenvalue density of the Dirac operator for $T\lesssim 1.2\,T_c$ 
has a small peak structure in the infrared favoring the form of $\rho(\lambda,m)\sim m^2\delta(\lambda)$, 
which can largely account for the origin of the axial anomaly. However, even in 
these studies a clear separation of the zero and near-zero
modes was not possible due to moderate residual chiral symmetry breaking effects
induced by the mixing of the left and right handed fermions along the finite fifth
dimension. On the other hand, another independent preliminary study with so-called optimal domain wall 
fermions and  physical pion mass reports the restoration of $U_A(1)$ above $T_c$~\cite{twqcd}.  
A study using overlap fermions restricted to the trivial topological sector of QCD 
on relatively small volumes also suggests that $U_A(1)$ is effectively restored at
$T\sim T_c$ \cite{jlqcd}. However, it is well known that simulations with fixed
topology are more sensitive to finite volume effects and, at present, it is
difficult to perform calculations with larger lattice volumes or for fluctuating
topology in the case of overlap fermions due to prohibitively large computational
costs.  

Measuring the underlying topology of the gauge fields on the lattice requires careful
analysis. The gauge fields are defined as links connecting the adjacent lattice
sites, which can be continuously deformed to unity. To study localized topological
structures one has to remove the ultraviolet fluctuations of the fields or approach
successively to the minimum of the classical action. The latter is done by cooling
the gauge configurations \cite{cooling}. The ultraviolet fluctuations can be reduced
using smearing \cite{smear}, which involves replacing each gauge link by an average
over the neighboring links.  It is then possible to measure the topological charge on
the lattice using the discretized version of the integrated $F\tilde F$ operator.
The most popularly used smearing technique, known as the hypercubic (HYP) smearing
\cite{hyp}, has been shown to provide a good estimate of the topological
susceptibility. However, successive smearing may lead to small instantons being
undetected and a change in the large scale structure of the gauge fields.
Alternatively, one can make use of the index theorem \cite{as}, which relates the
difference between the number of right and left-handed fermion zero modes to the
topological charge of the gauge fields. The advantage of this method is that it
naturally connects topological structures and the eigenmodes of the Dirac operator,
which are global quantities and depend on the gauge links of the entire lattice. 

In the present work we address the temperature dependence of $U_A(1)$ and
probable microscopic mechanisms responsible for its breaking in the
high temperature phase of QCD by studying the infrared eigenmodes of overlap
fermions \cite{neunar} on the background of dynamical (2+1) flavors of HISQ gauge field
configurations with nearly physical fermion masses and large volumes. This HISQ
discretization scheme has been used for extensive studies on QCD thermodynamics~\cite{hisqeos,hisqeos1} and 
has small discretization errors, resulting in the 
least taste symmetry breaking among all commonly used staggered fermion 
discretizations. Preliminary studies with HISQ fermions \cite{ding} also provide hints that in the 
continuum limit for two vanishingly small light quark masses, the QCD chiral transition may 
belong to the 3-dimensional $O(4)$ universality class. The issues about the lack of an 
index theorem of the HISQ is overcome by using the Overlap Dirac fermions to probe  
the topology of the HISQ gauge configurations.  Overlap Dirac fermions circumvent the Nielsen-Ninomiya No-go 
theorem \cite{nn} by sacrificing the ultra-locality criterion, preserve an exact 
chiral symmetry \cite{neunar} and an exact index theorem \cite{hln} and reproduce the 
correct anomaly \cite{adams1} even at nonzero lattice spacing. Employing the index 
theorem for the overlap fermions, the topological structures in SU(2) \cite{ehn} as well as 
SU(3) pure gauge theories \cite{ehn,gg} have been studied earlier. However, the relationship between chiral and $U_A(1)$ 
symmetry also cannot be addressed within the framework of pure gauge theory and the 
presence of light dynamical fermions is necessary to address this question. 
Furthermore, it is a priori not evident whether the same dilute instanton gas picture 
also applies for QCD with near-physical, light dynamical fermions as the presence of 
light fermions would lead to interactions between instantons and induce anomaly 
effects. 

This work is structured as follows: In Sec.~\ref{sc:details} we provide all
necessary computational details pertaining to this work.  In \ref{sc:chargedist} we
check the distribution of the topological charge measured by using the exact index
theorem of overlap fermions.  In \ref{sc:distribution} we present and discuss our
results on the eigenvalue distribution of overlap fermions on the dynamical HISQ
configurations.  The contribution of the low-lying eigenmodes towards $U_A(1)$
breaking is discussed in \ref{sc:fate_ua1}, while \ref{sc:evfit} contains our results about  the
functional form of the eigenvalue density and its implications for $U_A(1)$ breaking.
In \ref{sc:smear} we verify the robustness of the occurrence of near-zero modes.
From \ref{sc:topobj} to \ref{sc:localprop} we discuss various different properties of
the zero and near-zero eigenmodes.  Finally, in \ref{sc:conclusion} we summarize and
conclude this work. Preliminary results of this work were previously presented in
\cite{sayantan}.

\section{Computational details}
\label{sc:details}

The set of $(2+1)$-flavor HISQ configurations used in this work was generated by the
HotQCD collaboration \cite{hisqeos}. Two lattice sizes were used in this study,
$24^3\times6$ and $32^3\times 8$. The strange quark mass $m_s$ is set to its physical
value and the light quark mass in all these sets of configurations are chosen to be
$m_l=m_s/20$, which corresponds to a Goldstone pion mass of $m_\pi=160\,\mathrm{MeV}$
in the continuum. We studied 5 sets of configurations, two at $T\sim T_c$, two at
$T\sim1.2\,T_c$ and one at $T\sim1.5\,T_c$. Here, $T_c=154(9)$ MeV \cite{hisqeos} is the
chiral crossover temperature in the continuum limit.  Near $T_c$, in addition, we
studied configurations generated by the Bielefeld-BNL collaboration \cite{ding} with
lattice size $32^3\times 6$ and a light quark mass of $m_l=m_s/40$, which corresponds to
$m_\pi=110~$ MeV. This was to study whether the $U_A(1)$ breaking survives as the
chiral limit is approached.  We considered 90--160 configurations of each set,
typically separated by 100 trajectories, and computed the eigenvalues of the overlap
Dirac operator on them. The lattice sizes, strange to light quark mass ratio,
temperatures and relevant statistics are shown in Tab. \ref{table:stats}.

\begin{table}[h]
    \begin{ruledtabular}
        \begin{tabular}{ccccc}
            $N_\sigma^3\times N_\tau$ & $m_l/m_s$& $T$ [MeV] & $N$ & $N_\lambda$ \\
            \hline
            $24^3\times 6$&1/20& 162.3 & 120 & 200\\
            $32^3\times 6$ &1/40& 162.3 & 90 & 400\\
            $32^3\times 8$ &1/20& 165.6 &120 & 200\\ 
            $24^3\times 6$&1/20& 199.0 & 100 & 100 \\
            $32^3\times 8$ &1/20& 196.0 &100 & 100\\
            $32^3\times 8$ &1/20& 237.1 &160 & 50 \\
        \end{tabular} 
    \end{ruledtabular}
    \caption{ Lattice size ($N_\sigma^3\times N_\tau$), mass ratio ($m_l/m_s$),
    temperature ($T$), number of configurations ($N$) and number of eigenvalues that
    were computed per configuration ($N_\lambda$) for each ensemble.}
    \label{table:stats}
\end{table}

We probe the low-lying eigenmodes of these HISQ gauge ensembles through the use of
the massless overlap Dirac fermion operator
\begin{align}
\label{eq:overlapm}
D_\mathrm{ov} = M \left[ 1 + 
\gamma_5 \mathrm{sgn} \left[ \gamma_5 D_W(-M) \right] \right]
\;,
\end{align}
where $D_W$ is the standard Wilson Dirac operator with the parameter $0<M<2$.

For the implementation of the sign function in the overlap operator, we computed the
lowest 20 eigenvectors of $D_W^\dagger D_W$ using the Kalkreuter-Simma (KS) Ritz
algorithm~\cite{ks}. The sign function was computed for these low modes explicitly,
while for the higher modes it was approximated by a Zolotarev rational function. The
number of terms in the Zolotarev function was kept to be $15$. The overlap operator
satisfies the Ginsparg-Wilson(GW) relation with a deviation of no more than
$10^{-7}$ at low temperatures and $10^{-10}$ at high temperatures. The square of the
sign function deviated from identity by about $10^{-7}$--$10^{-9}$. 

For each temperature, 50 lowest eigenvalues of $\Dov^\dagger \Dov$
were computed using the KS algorithm.  The zero modes of $\Dov^\dagger
\Dov$ come with chiralities $\pm1$. The nonzero eigenvalues come in
degenerate pairs with chiralities having opposite signs but equal magnitudes, which
is usually different from unity. These features of the spectrum allow us to
distinguish between the near and exact zero modes within a few iterations of the KS
algorithm.  The KS algorithm was run until the relative error on the nonzero
eigenvalues of $\Dov^\dagger \Dov$ were estimated to be lower than
$10^{-4}$ on average and the separation between zero and nonzero modes was clearly
seen.  In most cases, the number of eigenvalues was later increased by computing the
eigenvalues of $P \Dov P$, where $P$ is the projection to righthanded or
lefthanded modes. This projected operator has the advantage that it only takes half
the time to be applied and each nonzero eigenvector can be related to a pair of
eigenvectors of $\Dov^\dagger \Dov$, further reducing computation time
and memory requirement. However, the algorithm becomes quite unstable if the subspace
that $P$ projects onto contains zero modes and can only be used on the opposite
chiralities after identifying the zero modes.

For some applications, the eigenvectors of $\Dov$ were required and not only those
of $\Dov^\dag \Dov$. Each degenerate pair of nonzero eigenvectors of the squared operator
spans a two-dimensional space that also contains two eigenvectors of $\Dov$, which
are related to each other by an application of $\gamma_5$ and have eigenvalues that
are complex conjugates of each other. These eigenvectors could be obtained by applying
an appropriate unitary transformation to each of the original pairs.

We also checked the optimal value of the parameter $M$ used in the construction of the overlap 
operator.  From partially quenched
studies it is known that for certain choices of $M$, the corresponding $D_W^\dagger
D_W$ can have very small eigenvalues, leading to the presence of spurious zero modes
in the overlap operator \cite{ovm}. We verified that for configurations without zero
modes the choice of $M$ did not affect the eigenvalues significantly within our
precision.  Moreover, for configurations with zero modes we chose $M$ such that the
sign function and the GW relation were determined with highest accuracy, ensuring the
best implementation of the overlap operator. Except for a few cases, especially near
$T_c$, we chose $M=1.8$.

For the $32^3\times 6$ lattice with a light quark mass of $m_l=m_s/40$, the gauge
configurations were rough and the convergence of the KS for $D_W^\dagger D_W$ was
slow, leading to imprecise estimates of the GW relation and the sign function.  In
this case, we did two levels of HYP smearing to smoothen out the ultraviolet
fluctuations. This enabled us to achieve a more precise estimation of the overlap
sign function, similar to the precision achieved for the other ensembles.  The
effects of the smearing are further discussed in Section \ref{sc:smear}. 

\section{Results}
\subsection{Topological charge distributions} \label{sc:chargedist}

\begin{figure}[t]
    \begin{center}
        \includegraphics[width=0.45\textwidth]{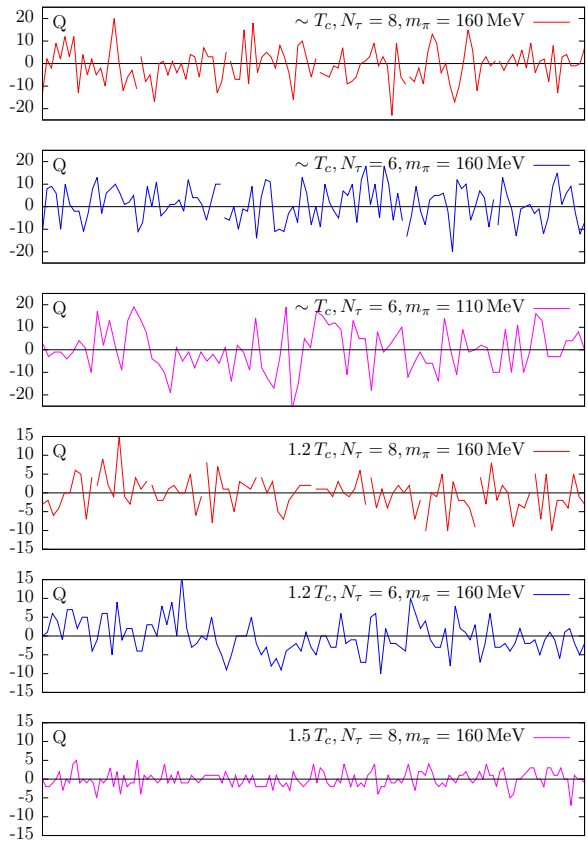}
        \caption{Time histories of the topological charge, calculated from the zero
        modes of the overlap Dirac operator, for the HISQ configurations at $T\sim
        T_c$, $T\sim1.2\,T_c$ and $T\sim1.5\,T_c$. Configurations belonging to the
        same production stream are connected with lines. }
        \label{toptrajec}
    \end{center}
\end{figure}

The topological charge $Q$ was measured by counting the number of 
zero-modes of the overlap operator and determining their chiralities,
\begin{align}
Q = n_+ - n_-~,
\end{align}
where $n_+$ ($n_-$) is the number of zero modes with chirality $+1$ ($-1$).
Since the underlying HISQ gauge configurations were
generated at zero strong CP violating angle $\theta$, at any temperature all the
topological sectors should be spanned.  However, 
the configurations may have been trapped in one topological sector and 
the autocorrelation times in such cases may be large.
To avoid autocorrelation effects we usually chose the configurations to be separated
by 100 Rational Hybrid Monte-Carlo trajectories.  The time histories of the
topological charge are shown in Fig.~\ref{toptrajec}. It is evident that the
autocorrelation effects are under control.  The distribution is ergodic enough and on
average $\langle Q \rangle\simeq0$, for all temperatures.  This gives us confidence
that the statistics is sufficient in our present study. Moreover, we also observe
many configurations with $|Q|\geq 1$ in all studied ensembles.  In fact, even at
$1.5\,T_c$ more than a third of the total number of configurations have $|Q|= 1$,
confirming the importance of the $Q\neq 0$ configurations. 

\subsection{Eigenvalue spectra} \label{sc:distribution}
In this section we show the eigenvalue density of the overlap operator at three temperatures
-- near $T_c$, at $1.2\,T_c$ and at a yet higher temperature of $1.5\,T_c$.
The overlap fermion matrix in Eq.~\eqref{eq:overlapm} is a normal matrix. In the
complex plane, its dimensionless eigenvalues, $\tilde \lambda$, lie on a circle
centered at $M$ and with a radius $M$, obeying $|\tilde\lambda-M|^2=M^2$. The
eigenvalues measured in our study lie on the circle very close to the origin with a very
small real part. Hence, we always plot the eigenvalue density as a function of
$\lambda$, where $a\lambda=\mathrm{Im}\tilde\lambda$. On the lattice, the eigenvalue
density is defined as
\begin{align}
    \label{eq:rholambda}
    a^3 \rho(\lambda)&=\frac{1}{N_\sigma^3 N_\tau}\sum_{i}\delta\left(a\lambda-a\lambda_i\right),
\end{align}
where the sum only includes values on the left part of the semicircle with
$\mathrm{Re}\tilde\lambda<M$, and excludes values near $\tilde\lambda=2M$ even though
their imaginary part would also be small. 

\begin{figure}[t]
    \begin{center}
        \includegraphics{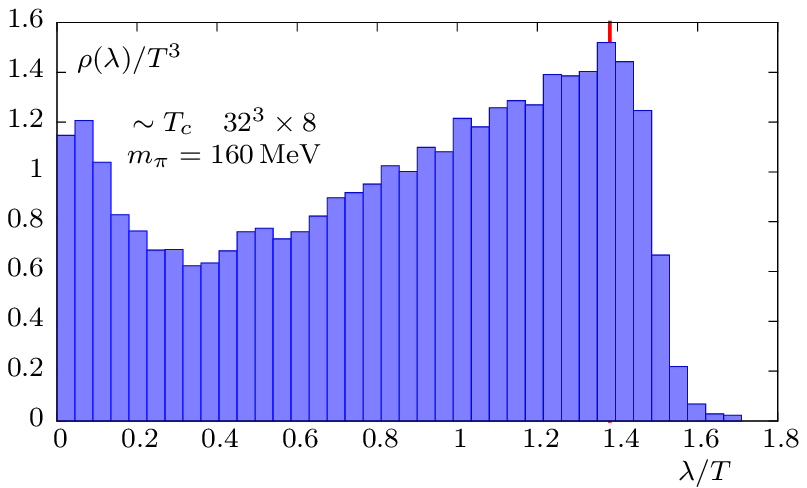}
        \includegraphics{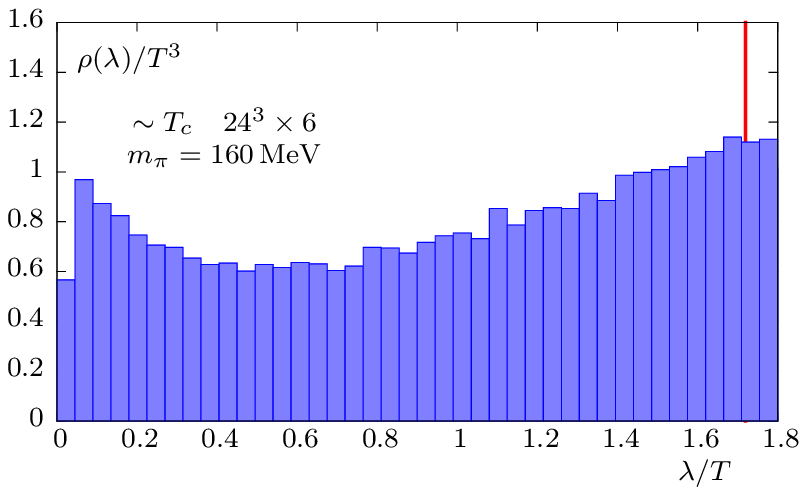}
        \includegraphics{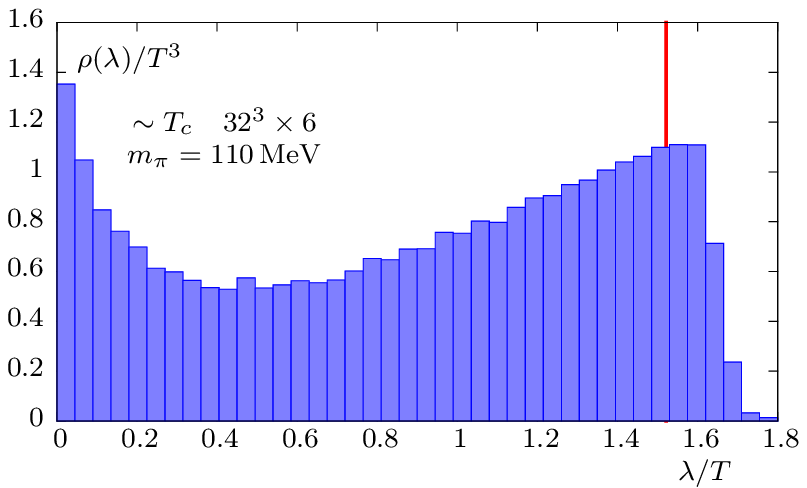}
        \caption{The eigenvalue density of the overlap operator on $32^3\times 8$,
        $24^3\times 6$ and $32^3\times 6$ HISQ configurations near $T_c$. Only
        nonzero modes are included. The vertical line denotes the range 
    of validity due to the finite number of computed eigenvalues.}
        \label{eigval1}
    \end{center}
\end{figure}

\begin{figure}[t]
    \begin{center}
        \includegraphics{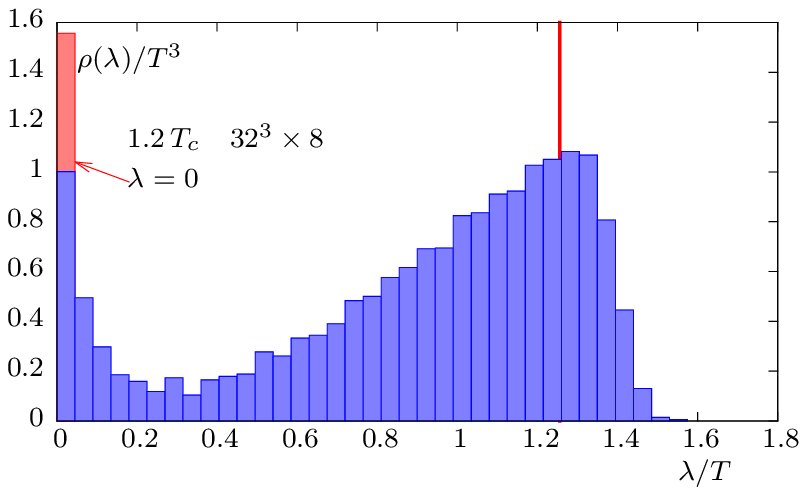}
        \includegraphics{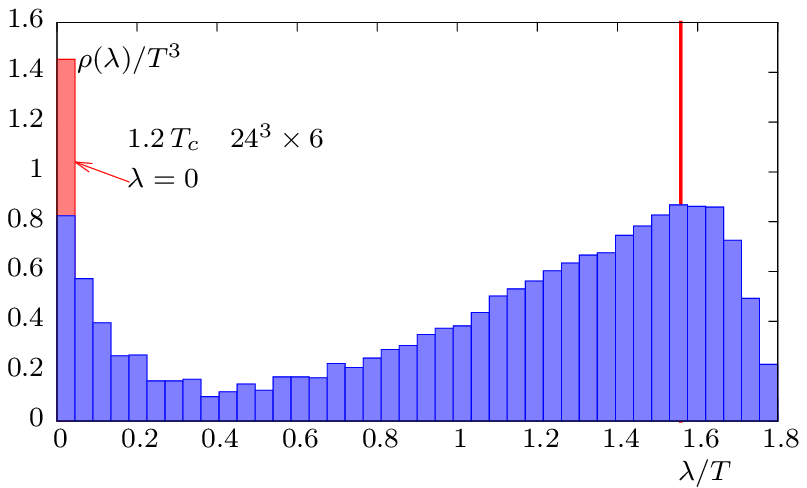}
        \caption{ The eigenvalue density for HISQ configurations using the overlap
        operator at $1.2~T_c$. The lattice sizes are $32^3\times8$ and $24^3\times
        6$, respectively. The red line marks the range of validity. }
        \label{eigval2}
    \end{center}
\end{figure}

\begin{figure}[t]
    \begin{center}
        \includegraphics{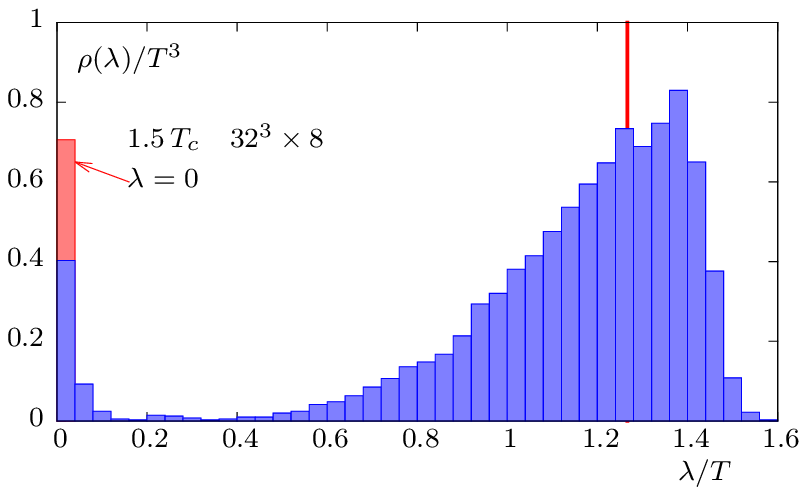}
        \includegraphics{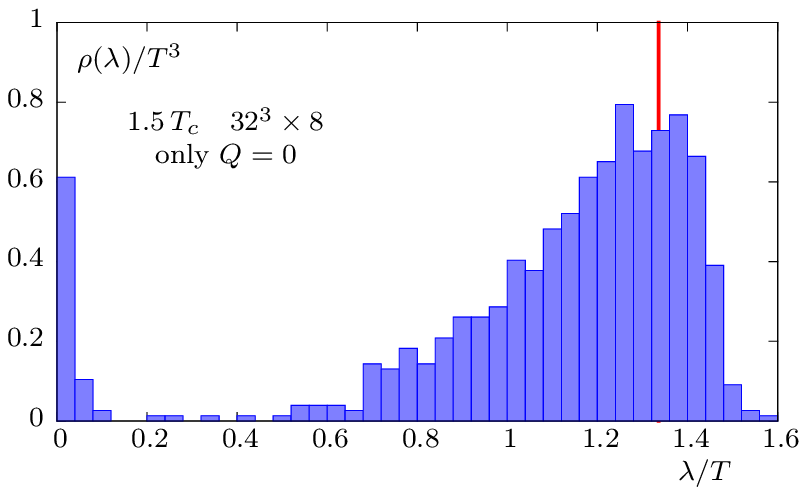}
        \caption{ The eigenvalue density for $32^3\times 8$ HISQ configurations using the
        overlap operator for 1.5 $T_c$ for all values of $Q$ and also separately for the
        $Q=0$ sector. The red line marks the range of validity. }
        \label{eigval3}
    \end{center}
\end{figure}

The eigenvalue distribution at three different temperatures is shown in
Figs.~\ref{eigval1}, \ref{eigval2} and \ref{eigval3}. The  spectrum is truncated
 at some large eigenvalue since we measure only a finite number of them. We
indicate the point beyond which the spectrum is not trustworthy anymore by a
vertical line in red in each of the plots. It is estimated by first taking the
highest computed eigenvalue of each configuration and then taking the minimum of
these values over all the analyzed configurations.

As we emphasized earlier, the KS algorithm allowed us to distinguish the zero modes
from the near-zero modes using the chirality properties of the corresponding
eigenvectors. In general, the eigenvalue distribution has three distinct features --
the zero mode peak, a near-zero mode accumulation and the bulk eigenvalue region.
Near $T_c$, 
the first bin contains a large contribution from zero modes which
are omitted in Fig.~\ref{eigval1} to focus on the infrared physics of only the
near-zero eigenvalues.  At this temperature, we do not observe any gap in the
infrared part of the eigenvalue spectrum. The near-zero modes and the bulk modes
appear to overlap significantly and the near-zero modes tend to develop a peak
towards the infrared region. This peak becomes sharper as the light sea quark mass is
lowered from $m_l=m_s/20$ to $m_l=m_s/40$ at fixed lattice spacing $1/6T$. It also
becomes sharper when we go to a finer lattice, from $N_\tau=6$ to $N_\tau=8$ at a
fixed pion mass of $160\,\mathrm{MeV}$. This trend suggests that the near-zero mode
accumulations will remain as the chiral and the continuum limits are approached. 

At temperatures $1.2\,T_c$ and $1.5\,T_c$, both the zero modes denoted by the red
bar and the near-zero and bulk modes are shown in Fig.~\ref{eigval2} and
Fig.~\ref{eigval3}. The separation between the near-zero mode
accumulation and the bulk eigenvalue region becomes even more evident with increasing
temperature. At $1.2\,T_c$, we study the eigenvalue spectrum at two different
lattice spacings to estimate whether the infrared part of the spectrum is strongly
affected by the lattice cutoff effects at higher temperatures. Keeping the physical
bin size the same in units of $\lambda/T$ for comparison, we observe that the
infrared region of the eigenvalue density remains  practically unchanged when the lattice spacing
goes from $1/6T$ to $1/8T$ at a fixed temperature $T$. This gives us confidence that
the near-zero modes are not due to dislocations of the gauge fields.  A more detailed
study about the lattice artifacts is given in Sec.~\ref{sc:evfit}.

The number of zero and near-zero modes both decrease as the temperature is increased
to $1.5\,T_c$ as shown in Fig.~\ref{eigval3}. There is a small peak of near-zero
modes, while the number of bulk eigenvalues starts to rise very slowly and only gives a significant
contribution beyond $\lambda_0 \simeq 0.4\,T$.
This is reminiscent of some kind of band edge separating the two
different regimes of eigenvalues, which is studied in detail in
Sec.~\ref{sc:localprop}.  Even at this temperature  we do not observe a gap in the
infrared sector of the eigenvalue spectrum. The presence of these near-zero modes is
not due to the fact that we are also sampling configurations belonging to nonzero
topological sectors in our study.  This is evident from the lower panel of Fig.~\ref{eigval3},
where we show the
eigenvalue distribution of only those configurations with topological charge $Q=0$ at
$1.5\,T_c$. The presence of near-zero modes is also observed for
this particular subset of configurations. 

\subsection{Near-zero eigenmodes and axial symmetry breaking} \label{sc:fate_ua1}
\newcommand{\ddm}{\partial_m}
As introduced in Sec.~\ref{sc:intro}, $\omega\equiv\chi_\pi-\chi_\delta$ as
defined in Eq.~\eqref{eqn:chipd} is a measure that quantifies $U_A(1)$ breaking. 
Through a chiral Ward identity it can also be obtained 
\begin{align}
\label{eqn:omega}
 \omega&=\frac{\langle\bar\psi\psi\rangle}{m}-\chi_{\text{conn}}
\end{align}
from the chiral condensate 
$\left<\bar\psi\psi\right> = \frac{T}{V}\left<\mathrm{tr}\left(D_m^{-1}\ddm{D_m}\right)\right>$
and the connected chiral susceptibility
$\chi_\text{conn}=\frac{T}{V}\left<\ddm{\mathrm{tr}\left(D_m^{-1}\ddm{D_m}\right)}\right>$,
where $D_m=\Dov (1-am/2M)+am$ is the Dirac operator for overlap quarks with
a (valence) quark mass $m$.

Thus, in terms of the eigenvalues of the overlap operator
\begin{align}
    \label{eqn:chipi}
    a^2\omega&=
    \frac{1}{N_\sigma^3 N_\tau}
    \left[
        \frac{\langle\vert Q\vert\rangle}{(am)^2}\right.\\\nonumber
        &+\left.
        \left\langle\sum_{\tilde\lambda\neq 0}
        \frac{2(am)^2(4M^2-|\tilde\lambda|^2)^2}
        {\left[|\tilde\lambda|^2(4M^2-(am)^2)+4 (am)^2 M^2\right]^2}
    \right\rangle\right]~.
\end{align} 
The first term is the 
contribution from the zero modes which  vanishes in the thermodynamic
limit.

\begin{figure}[t]
    \centering
    \includegraphics{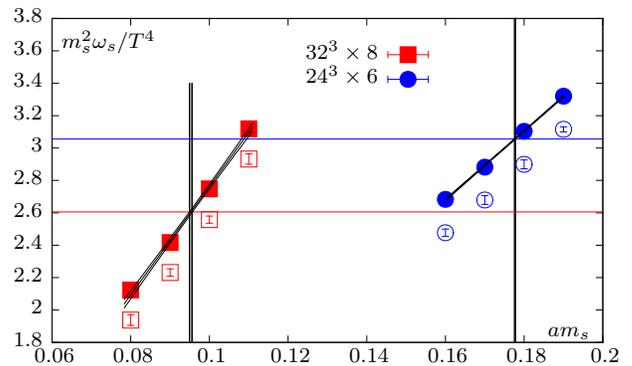}
    \caption{The tuning of the strange valence quark mass for overlap fermions on
    $N_\tau=6$ (blue) and $N_\tau=8$ (red) lattices near $T_c$. The horizontal lines mark the
    results for $m_s^2\omega_s/T^4$ (cf. Eq.~\eqref{eqn:omega}) independently 
    obtained for the HISQ sea fermions.
    Filled (empty) points denote the overlap result with (without) the zero mode
    contribution.}
    \label{fig:cpcdTc}
\end{figure}

Thus, having determined the low-lying eigenvalues of the overlap operator,
$\omega$ can be computed from them.  However, to explore the physics of
the underlying HISQ configurations the overlap valence quark mass 
that enters Eq.~\eqref{eqn:chipi}  has to be
tuned against some physical quantity measured on the same gauge configurations. In
the present work we adopt a very simple strategy: we roughly tune the strange
valence quark mass, $m_s$, and then study $\omega$ as a function of the 
light valence quark mass within a range of $m_l=m_s/20$ to $m_l=m_s/2$. To tune the
strange valence quark mass we use the renormalized difference 
of the pseudo-scalar ($\eta_{s\bar s}$) and the (connected) scalar susceptibilities 
in the strange quark sector, $m_s^2\omega_s/T^4$.
We calculated this renormalized
quantity using strange valence overlap fermions and matched it with the same 
quantity calculated independently for the strange HISQ sea quarks.
For the overlap fermions this quantity was estimated in two
parts. We first  calculated this quantity from the predetermined
low-lying eigenvalues using Eq.~\eqref{eqn:chipi}, 
and then added the contribution of higher eigenvalues by
performing inversions on the eigenspace orthogonal to the low-lying eigenmodes using 
random source vectors. The result of the strange valence  quark mass tuning near
$T_c$ is shown in Fig.~\ref{fig:cpcdTc}. 
In order to monitor finite volume effects,
we considered the exact zero modes separately and do not observe a significant contribution
from them to this quantity. The tuned masses obtained with and without the
zero modes differ by about $5\%$.

\begin{figure}[t]
    \centering
    \includegraphics{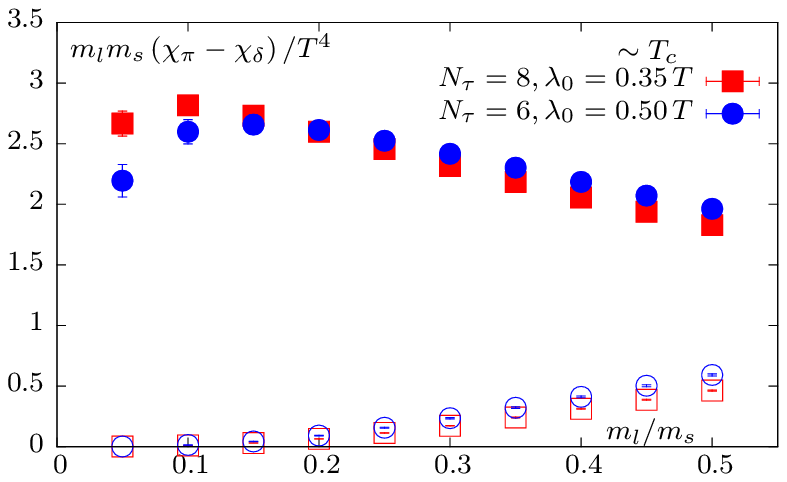}
    \includegraphics{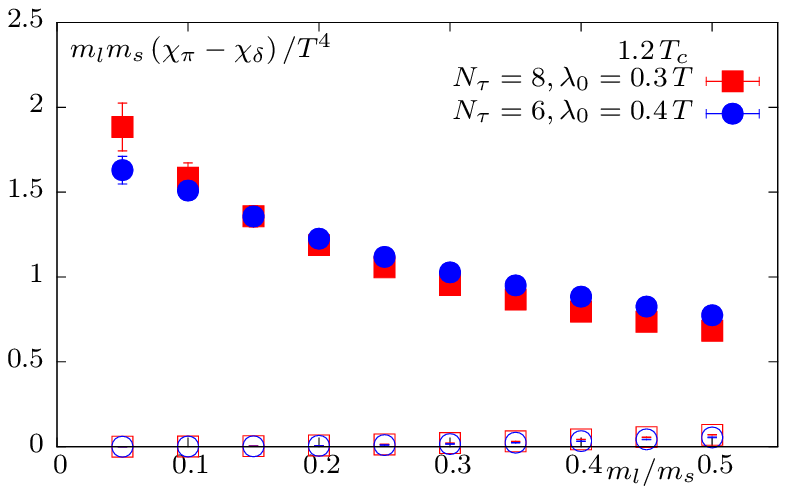}
    \caption{A renormalized measure of $U_A(1)$ breaking for a range of
    valence light quark masses, $m_s/20\leq m_l \leq m_s/2$ for ensembles with different
    $N_\tau$ at $T\sim T_c$ and $T\sim1.2\,T_c$. The filled points denote the
    contribution from near-zero modes ($\lambda<\lambda_0$), while the empty points
    were calculated only from the bulk modes ($\lambda>\lambda_0$). }
    \label{fig:cpcdlight}
\end{figure}

In Fig.~\ref{fig:cpcdlight}, we show a renormalized measure of $U_A(1)$ breaking,
namely $m_lm_s\omega/T^4$, for a range of the 
light valence quark mass between $m_s/20$ and $m_s/2$.  Assuming a Breit-Wigner distribution
for the near-zero mode peak to model the $\delta(\lambda)$ like distribution discussed in the
introduction, i.e.  $\rho(\lambda)/T^3=\rho_0 A/(A^2+\lambda^2)$, the
contributions of the near-zero modes in this renormalized measure of $U_A(1)$
breaking can be characterized as 
\begin{align}
    \frac{m_lm_s}{T^4}\omega &\propto
    \rho_0 m_s\frac{A+2m_l}{(A+m_l)^2}~.
\end{align}
In our partially quenched setup, where only the valence light quark mass $m_l$ is
varied, this quantity has a finite value in the chiral limit $m_l\to0$. In a full
dynamical setup, its behavior will be governed by the  dependence of $A$ and
$\rho_0$ on the light sea quark mass. 
The simplest case of both $A$ and $\rho_0$ being proportional to the light sea quark mass,
which is compatible with the trends discussed in Sec.~\ref{sc:evfit}, will also give a finite value
and should be approximated in a partially quenched study.  For light 
valence quark masses near to or smaller than the smallest near-zero eigenvalues, the computed quantity
will approach zero, which is visible for the lowest masses near $T_c$.
This is a finite volume effect in the sense that a larger volume will 
sample more eigenvalues in the near-zero mode region, which
reduces the magnitude of the smallest eigenvalue and pushes this effect towards zero.
Beyond that, we observe a smooth dependence on $m_l/m_s$ that is compatible with the Breit-Wigner
ansatz and which is independent of the lattice spacing.
It is also evident from Fig.~\ref{fig:cpcdlight} that the contribution of the near-zero modes
to $\omega$ is substantially larger than that from the bulk modes.
When going from $T\sim T_c$ to $T\sim1.2\,T_c$, this quantity
does not decrease significantly, supporting our conclusion that $U_A(1)$ is not
effectively restored simultaneously with the chiral symmetry.

\subsection{The functional form of the eigenspectra} \label{sc:evfit}

In order to understand the general functional form of the eigenvalue density and in
particular the near-zero region, we make a fit ansatz consisting of a Breit-Wigner
peak for the near-zero modes and a polynomial behavior for the bulk part of the
spectrum of the form,
\begin{align}
\label{eqn:evfit}
\frac{\rho(\lambda)}{T^3} &= \frac{\rho_0A}{A^2+\lambda^2}+c\lambda^\alpha
~.
\end{align}
We address three issues in this section. Firstly, a general idea about the dependence
of the near-zero mode peak on the sea quark mass is necessary to understand what
happens in the chiral limit. Secondly, it is important to check the dependence of the
near-zero modes on the lattice cutoff to establish that these are physical and not 
mere lattice artifacts.
Finally, the leading exponent that characterizes the rise of the bulk also provides
information regarding the restoration of $U_A(1)$, hence its dependence on the
temperature and lattice cutoff needs to be studied. 

\begin{figure}
    \begin{center}
        \includegraphics[width=0.45\textwidth]{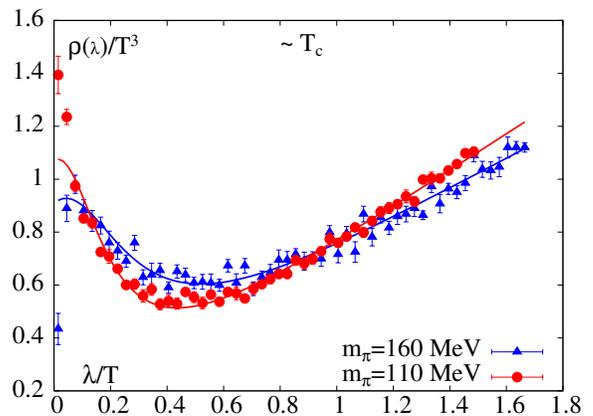}
        \caption{Eigenvalue distribution at $T\sim T_c$ for $N_\tau=6$ and two
            different sea quark masses compared to fits with Eq.~\eqref{eqn:evfit}.}
        \label{fig:eigval5}
    \end{center}
\end{figure}

The fit to the eigenvalue spectrum near $T_c$ for different sea quark masses is shown
in Fig. \ref{fig:eigval5}. The error bars for each bin have been determined by a
jackknife procedure over the set of gauge configurations.  The parameter $A$,
characterizing the width of the near-zero mode peak, falls from $0.35(4)\,T$ to
$0.151(7)\,T$ when going from $m_l=m_s/20$ to $m_s/40$ and the prefactor $\rho_0$, indirectly
controlling the height of the peak, also goes down from $0.28(3)\,T$ to $0.191(7)\,T$.
This generally supports the picture of a delta function like peak forming in the
chiral limit, with a decreasing peak height. However, as far as one can tell from two data
points, the dependence of $\rho_0$ does not seem to be quadratic in the light sea quark
mass near $T_c$, arguing against the dilute instanton gas picture as a good description of QCD 
at this temperature.

\begin{figure}
    \begin{center}
     \includegraphics[width=0.45\textwidth]{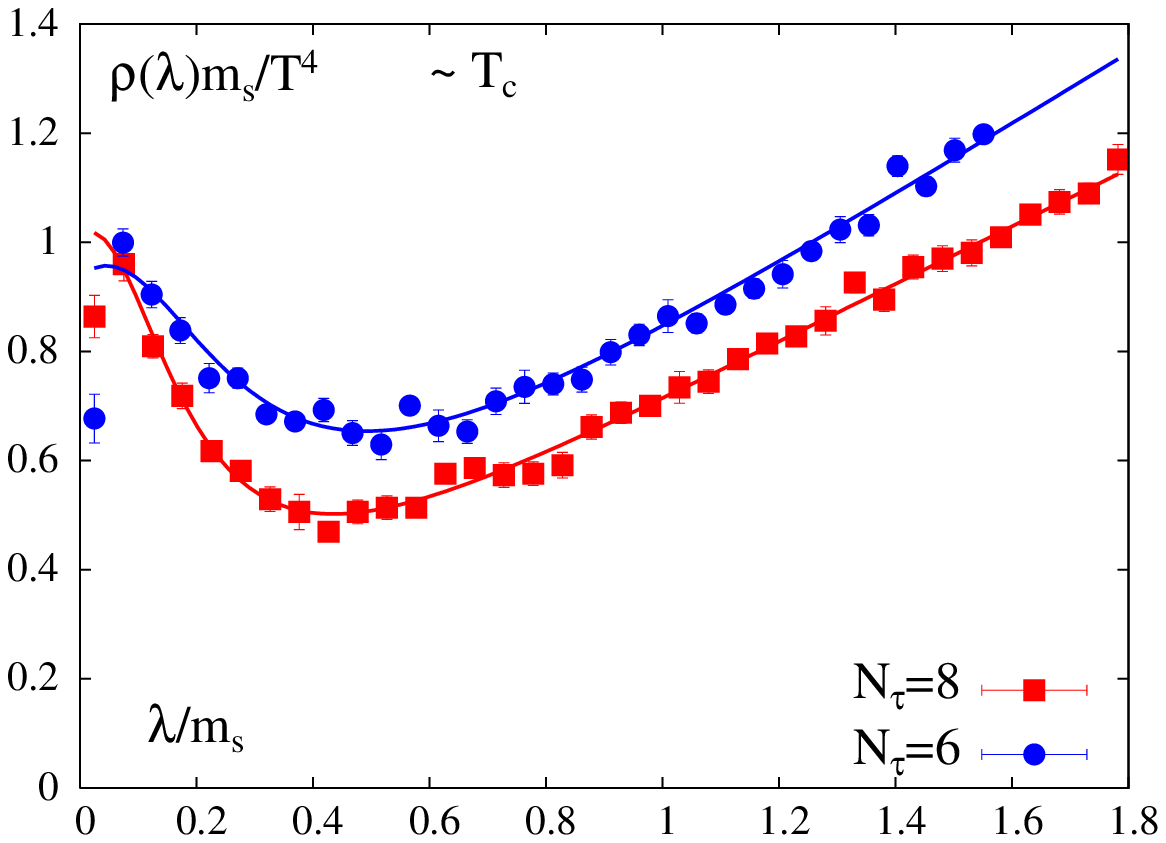}
      \includegraphics[width=0.45\textwidth]{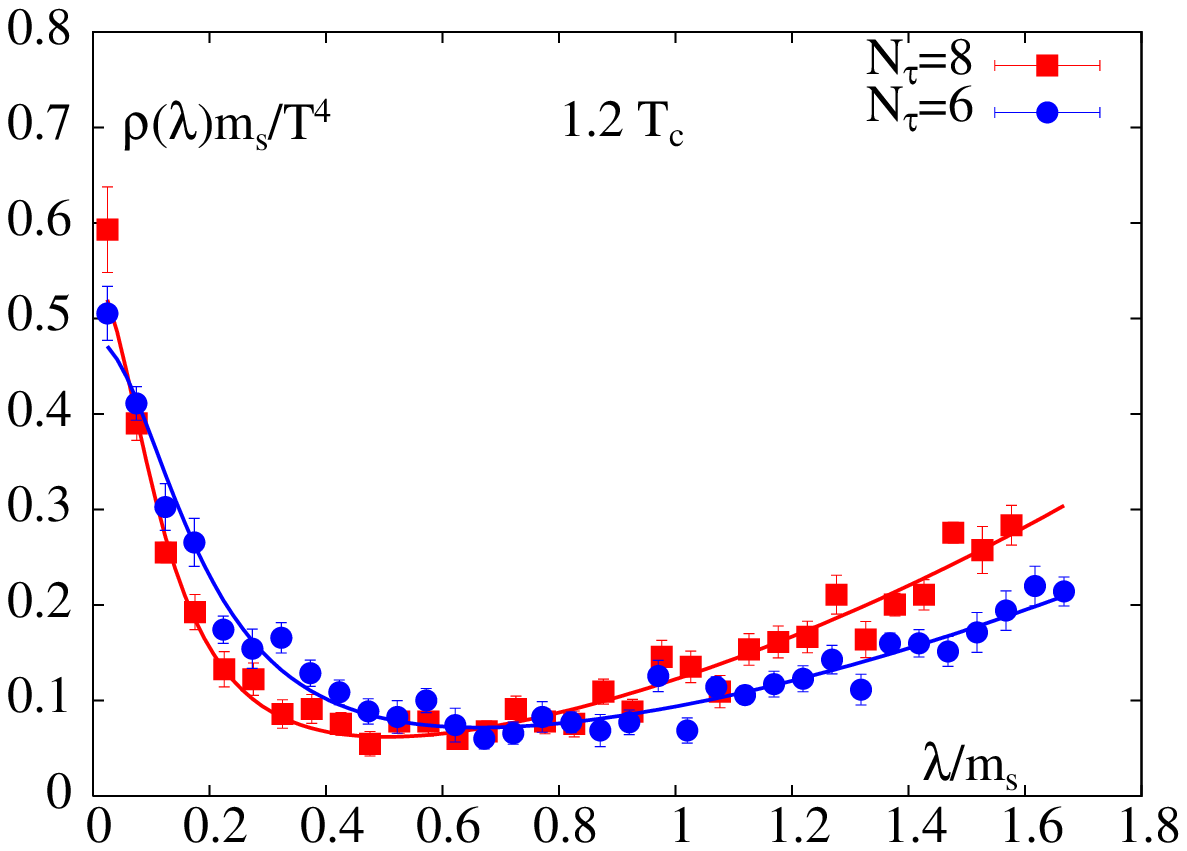}
        \caption{The renormalized eigenvalue spectra near $T_c$ (top) and at
        $1.2\,T_c$ (bottom), both for two different lattice spacings. The lines indicate
        fits to Eq.~\eqref{eqn:evfit} with $\lambda\rightarrow\lambda/m_s$ and
        $\rho(\lambda)/T^3\rightarrow m_s\rho(\lambda)/T^4$.}
        \label{fig:eigval6}
    \end{center}
\end{figure}

To examine cutoff effects one needs to compare a 
renormalized version of the eigenvalue density at different lattice spacings. One way to
renormalize the eigenvalues is to scale them by the previously tuned strange quark
mass. The corresponding renormalized eigenvalue density is then $m_s \rho(\lambda)$
since it leaves the quantity $m_s\langle\bar \psi\psi\rangle$ unchanged under
renormalization. We therefore take the same ansatz as in Eq.~\eqref{eqn:evfit} with
$\lambda\rightarrow\lambda/m_s $ and the density replaced by its renormalized
definition $m_s\rho(\lambda)/T^4$. Fits to the renormalized spectrum in dimensionless
units at two different temperatures, $T_c$ and $1.2~T_c$, are shown in Fig. \ref{fig:eigval6}.
Taking a closer look at the near-zero mode peak, it is evident that the accumulation of
near-zero modes is almost independent of the lattice spacing and is unlikely to be
just a lattice artifact.  

Finally, we discuss the exponent $\alpha$ of the characteristic $\lambda^\alpha$ rise
of the bulk eigenvalues. As mentioned in Sec.~\ref{sc:intro}, under the assumption
of analyticity of the eigenvalue density in $m^2$, detailed analytical calculations
based on up to 4-point chiral Ward identities show~\cite{aoki} that in the chiral
symmetric phase of QCD the leading $\lambda$ dependence should be similar to that for
the free theory, i.e. $\lim_{m\to0} \rho(\lambda,m) \sim
\lambda^3$. In such a case, the effect of $U_A(1)$ breaking should be invisible in at
least up to 6-point correlation functions. In light of this, it is
interesting to characterize the rise of the bulk eigenvalues. As shown in Fig.~\ref{fig:eigval5},
near $T_c$ the rise of the bulk eigenvalues for our two $N_\tau=6$
lattices with quark masses $m_l=m_s/20$ and $m_l=m_s/40$ is described by
the exponents  $\alpha=0.92(5)$ and $\alpha=0.98(4)$, respectively.  A fit to the
renormalized eigenvalue spectrum near $T_c$ further yields 
$0.86(2)$ for $N_\tau=8$, see Fig.~\ref{fig:eigval6}. Thus, near $T_c$ a linear
rise of bulk eigenvalues is favored for both lattice spacings and quark masses.
Interestingly, chiral perturbation theory~\cite{smilga} and the ILM~\cite{verbaarschot}
show that a contribution to $\rho(\lambda)$ linear in $\lambda$,
which results in a non vanishing connected susceptibility $\chi_\text{conn}$, is absent for two light
flavors and only present for $N_f>2$. On the other hand, for staggered fermions away from the 
continuum limit, taste violations lead to $\chi_\text{conn}\neq 0$ also for $N_f=2$~\cite{prelovsek}. 

\begin{figure}
    \begin{center}
        \includegraphics[width=0.45\textwidth]{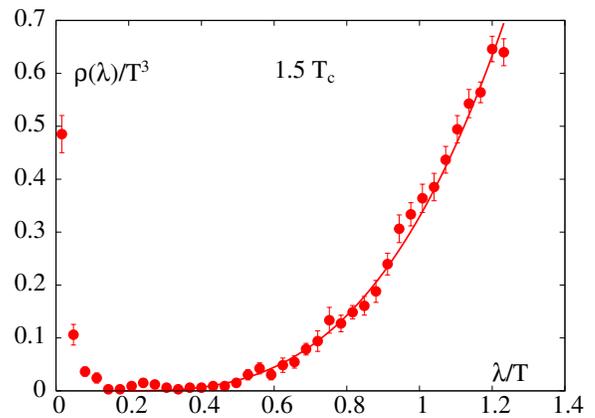}
        \caption{Eigenvalue distribution at $1.5\,T_c$ together with
        $a\left(\lambda-\lambda_0\right)^\alpha$ fit to the bulk eigenmodes.}
        \label{fig:eigval7}
    \end{center}
\end{figure}

At $1.2\,T_c$ the rise of the bulk eigenvalues has a different exponent. From 
 Fig.~\ref{fig:eigval6} it is evident that $\alpha\simeq2$, independent of the lattice
spacing. Note that a bulk eigenvalue density rising quadratically with $\lambda$
does not contribute to $\omega$.
A similar linear behavior for $T\sim T_c$ and a quadratic rise for
$T\sim1.2\,T_c$ of the bulk eigenvalues was also observed in the previous study
with domain wall fermions \cite{dw2} with a heavier pion mass, corroborating that
the bulk rise is independent of the sea quark mass. On the other hand, the characteristic 
free theory like cubic rise of the bulk eigenvalue density was only observed at 
$1.5\,T_c$ as shown in Fig.~\ref{fig:eigval7}.  At $1.5\,T_c$,
the near-zero mode peak reduces significantly and the separation between the bulk
and the near-zero modes is distinctly visible.  Our fit ansatz for the bulk is
modified accordingly as $(\lambda-\lambda_0)^\alpha$ to represent this feature, which
gives a smaller $\chi^2$ per degrees of freedom than the original ansatz in
Eq.~\eqref{eqn:evfit}. The parameter $\alpha$ and the goodness of fit at different
temperatures are compiled in Tab.~\ref{tab:fitparam}. 

\begin{table}[h]
    \begin{ruledtabular}
        \begin{tabular}{ccccc}
           $N_\sigma^3\times N_\tau$ & $m_l/m_s$& $T$ [MeV] & $\alpha$ &
           $\chi^2/\text{dof}$ \\ \hline
            $24^3\times 6$&1/20& 162.3 & 0.92(5) & 1.32\\
            $32^3\times 6$ &1/40& 162.3 & 0.98(4) & 1.84\\
            $32^3\times 8$ &1/20& 165.6 & 0.86(2) & 0.92\\ 
            $24^3\times 6$&1/20& 199.0 & 1.9(2) & 1.16 \\
            $32^3\times 8$ &1/20& 196.0 &1.9(1) & 1.21\\
            $32^3\times 8$ &1/20& 237.1 & 3.0(4) & 1.30\\
        \end{tabular} 
    \end{ruledtabular}
    \caption{Lattice size ($N_\sigma^3\times N_\tau$), mass ratio ($m_l/m_s$),
    temperature ($T$), the exponent $\alpha$ characterizing the $\lambda^\alpha$ rise
    of the bulk eigenvalues $\lambda$ and the goodness of the fits performed on the
    eigenvalue distribution. }
    \label{tab:fitparam}
\end{table}

\subsection{Robustness of the zero and near-zero modes} \label{sc:smear}

Detecting topological objects with fermion zero modes has the advantage that by 
construction the zero modes depend on all the gauge links distributed on
a lattice.  Still, if the underlying gauge fields are not smooth enough, the method
might be hampered by the presence of unphysical fermion modes localized on structures called
dislocations, which typically have a smaller classical action than instantons. These
are lattice artifacts, i.e. effects of finite lattice spacing, and should disappear
as the continuum limit is approached. It is therefore important to  make sure that
the observed infrared fermion modes are physical and do not solely arise as 
lattice artifacts.  If the zero and near-zero eigenvalues are entirely due to the presence 
of dislocations, they are expected to disappear as the gauge fields are smoothed using smearing
techniques. To check this we performed HYP smearing on the $32^3\times8$
configurations with $m_l/m_s=1/20$ at $1.5\,T_c$.  

Such smoothing methods are mandatory if one wants to compute
the topological charge by means of a discretized version of its field theoretic
definition
\begin{align}
    Q &= \frac{1}{32\pi^2}\int\!\mathrm{d}^4x F_{\mu\nu}^a \tilde F_{\mu\nu}^a~.
\end{align}
The optimal number of smearing levels is usually chosen such that the topological
charge measured  this way on the smeared configurations has a value close to an integer.
On our present lattices it turned out that $10$ levels of HYP smearing were sufficient to give an integer
value of the $F\tilde F$ operator summed over the whole lattice at $1.5\,T_c$.  
On all of these smoothed configurations we could verify that the value of $Q$ obtained from $F\tilde F$
matched exactly with the topological charge obtained by counting the zero modes of the 
overlap operator after smearing.

\begin{figure}[t]
    \begin{center}
        \includegraphics[width=0.4\textwidth]{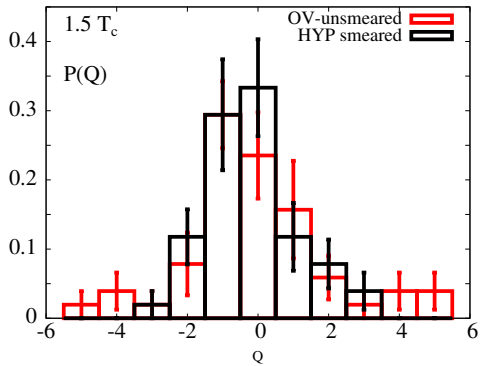}
        \caption{The distribution of the topological charge, $Q$, at $1.5\,T_c$ for
        $32^3\times8$ HISQ configurations measured from the zero modes of the overlap
        operator and also from the gluonic operator $F\tilde F$ on the same gauge
        configurations after $10$ levels of HYP smearing.}
        \label{tophistogram}
    \end{center}
\end{figure}

In Fig.~\ref{tophistogram}, we compare the histogram for the 
topological charge measured by counting the fermion zero modes of unsmeared configurations 
with that measured using the purely gluonic observable $F\tilde F$ on 
the same configurations after smearing.  
Apparently, some zero modes disappear in the course of the smearing process; yet, the comparison
indicates that the fermionic zero modes for these configurations do not arise only due to gauge
field dislocations and are reflecting continuum physics. 

\begin{figure}[t]
    \begin{center}
        \includegraphics{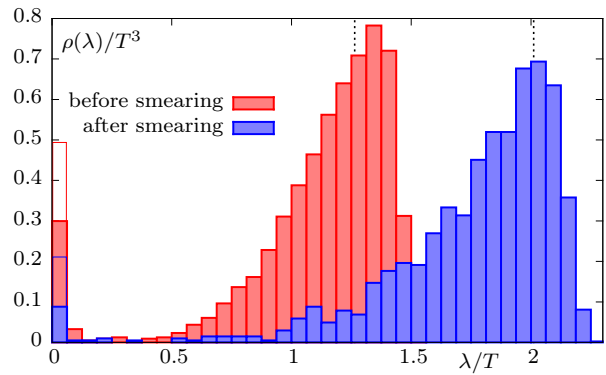}
        \caption{The eigenvalue density at $1.5\,T_c$ and $N_\tau=8$ before and after 10 steps of HYP smearing.
            The empty boxes are obtained when including zero modes and the dashed lines mark the ranges of validity.
        }
        \label{hypcomp}
    \end{center}
\end{figure}

A similar behavior is observed for the near-zero modes. In Fig.~\ref{hypcomp}, the 
comparison of the eigenvalue spectrum of the overlap operator on the original 
unsmeared configurations with the smeared ones reveals that the number of near-zero 
modes is somewhat reduced by smearing but that the near-zero mode accumulation is
still present even after a substantial amount of smearing, indicating that these are
not mere lattice artifacts like dislocations. We also found that the typical
eigenvectors associated with the near-zero modes on a smeared configuration appear
to be slightly less localized compared to the unsmeared case, suggesting that the
reduction of the near-zero modes may be caused by the loss of small instantons due to
smearing. 

Finally, as already shown in Fig.~\ref{fig:eigval6},  near $T_c$ as well as at $1.2\,T_c$ 
the  comparison of the renormalized eigenvalue spectra at two different lattice spacings,
$1/6T$ and $1/8T$, indicates that the
near-zero mode accumulation remains nearly unchanged as the lattice spacing is
reduced. Thus, it is unlikely that the near-zero mode accumulation arises primarily due
to lattice artifacts. 

\subsection{Profiles of the zero and near-zero modes at high temperature} \label{sc:topobj}

The fermion zero mode associated with an instanton is, at $T=0$, localized in the
region occupied by the instanton.
At nonzero temperature however, the compactification of Euclidean time
leads to periodic copies of instantons. Such classical finite action solutions of the gauge 
fields on the manifold $R^3\times S^1$ are known as calorons.  
When the instanton size is much smaller than $1/T$, the copies 
do not feel the effect of the neighbors and behave like zero
temperature instantons.  However, the overlap between the instanton copies may be
larger and their sizes become comparable to 
$1/T$. Explicit solutions are known for trivial \cite{hshep} as well as nontrivial~\cite{kvbaal,lu}
holonomy. In the case of trivial holonomy it has been observed that the caloron turns
into a magnetic monopole \cite{rossi} in pure gauge theory.  Calorons with 
nontrivial holonomy have more interesting features 
with monopole substructures \cite{mp4}. However, in our studies we have not attempted 
to take a detailed look at the monopoles.

\begin{figure}[t]
    \begin{center}
        \includegraphics[width=0.4\textwidth]{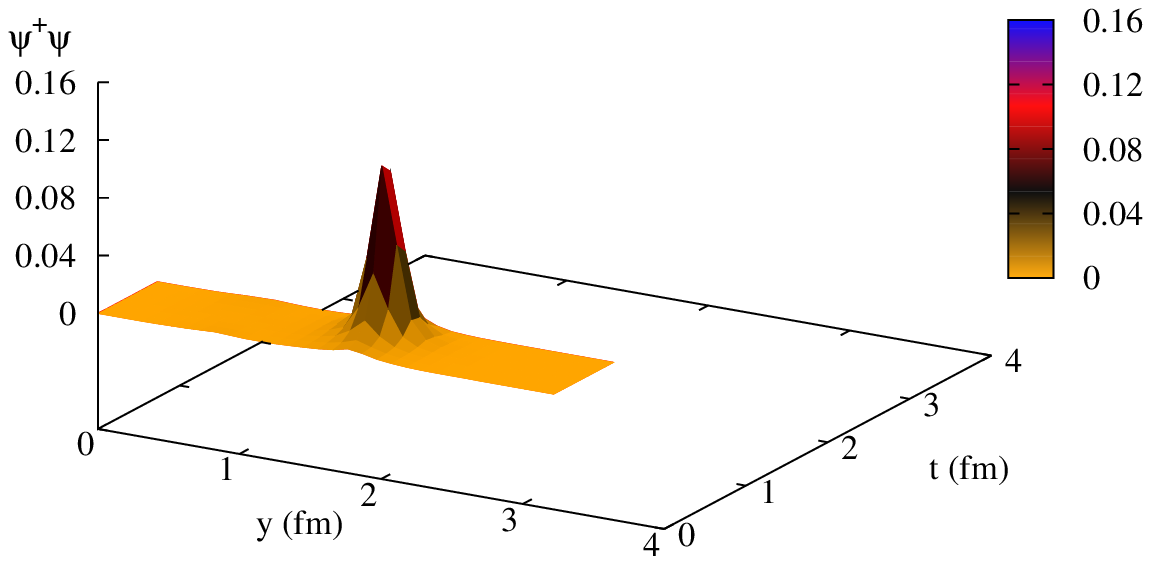}
        \includegraphics[width=0.4\textwidth]{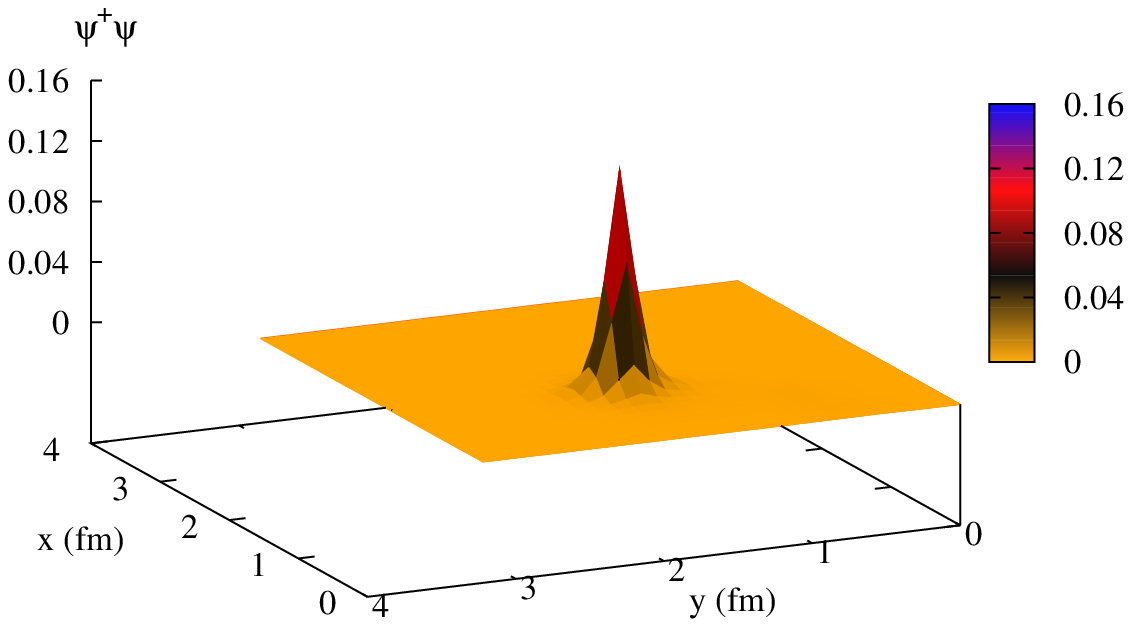}
        \caption{Space-time profile of a zero mode at $1.5\,T_c$ for a typical
        gauge configuration with $Q=1$.}
        \label{xyt1}
    \end{center}
\end{figure}

\begin{figure}[t]
    \begin{center}
        \includegraphics[width=0.4\textwidth]{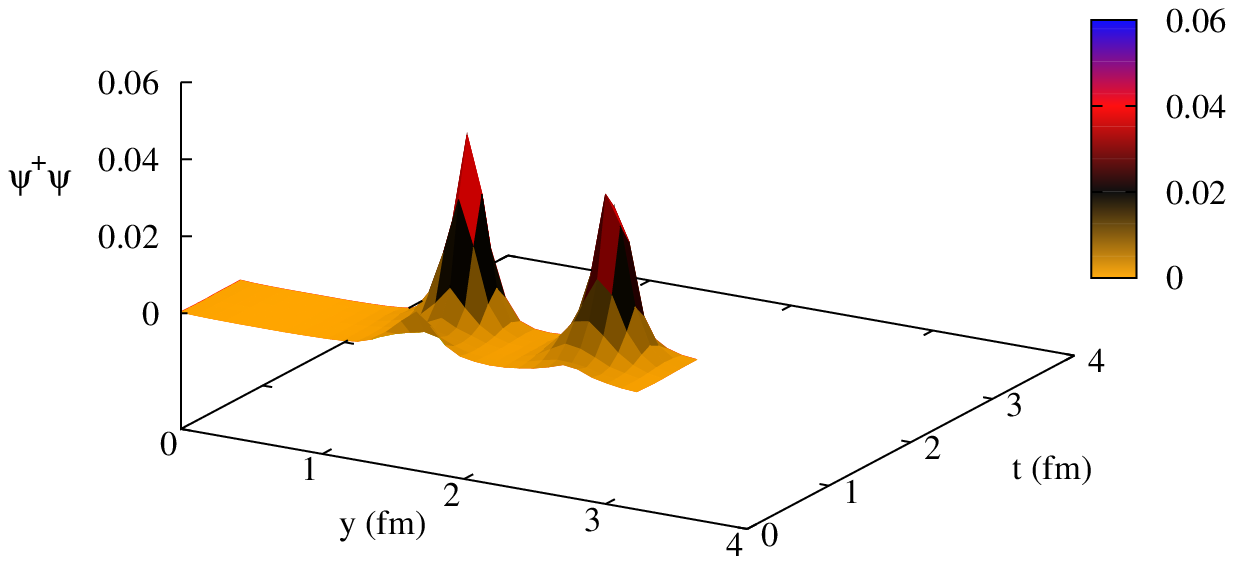}
        \includegraphics[width=0.4\textwidth]{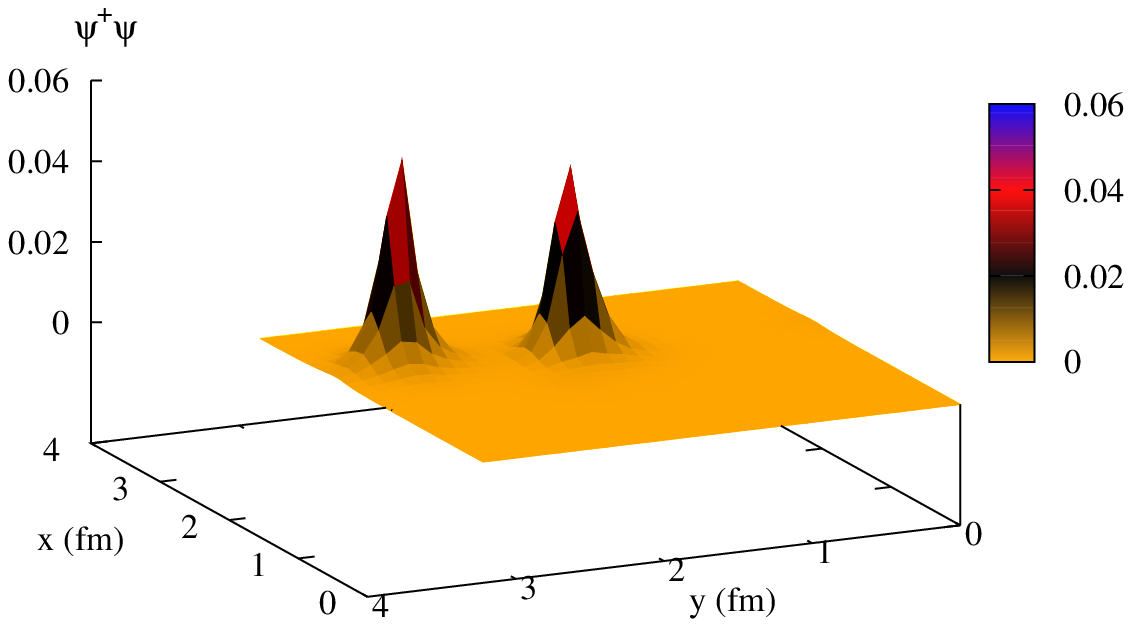}
        \caption{Space-time profile of a near-zero mode at $1.5\,T_c$ for the same
        gauge configuration depicted in Fig.~\ref{xyt1}. }
        \label{xyt2}
    \end{center}
\end{figure}

In order to gain more inside into the
structure of the infrared modes in QCD, we looked at the profiles of the zero and the
near-zero modes at $1.5\,T_c$.  Examples representing the majority of configurations
with topological charge $|Q|=1$ are shown in Figs.~\ref{xyt1} and \ref{xyt2}. 
In these figures the density of the wavefunctions, $\psi^\dagger(x)\psi(x)$, is measured along two 
spacetime directions, summing over the other two directions and the internal
degrees of freedom (color and spin). We observe that the zero modes are localized along
the spatial as well as the temporal directions. 
In a few cases, the width of the zero modes in the compact temporal direction is somewhat larger
and the density profile represents the overlap between the nearest copies.

\begin{figure}[t]
    \begin{center}
        \includegraphics[width=0.4\textwidth]{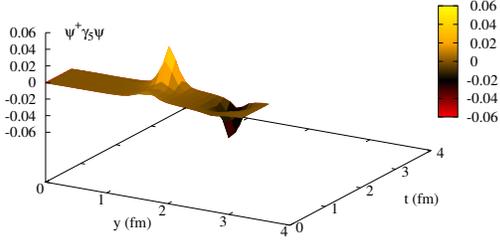}
        \includegraphics[width=0.4\textwidth]{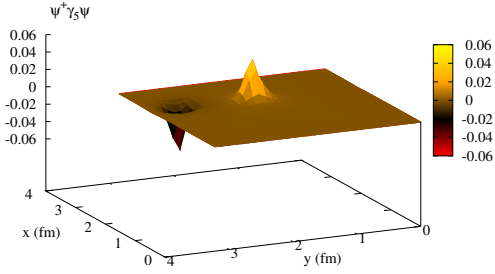}
        \caption{Space-time profile of chirality of the same near-zero mode depicted
        in Fig.~\ref{xyt2} at $1.5\,T_c$.}
        \label{chiral}
    \end{center}
\end{figure}

The near-zero modes typically exhibit a two-peak structure, in the density 
as well as in the chirality ($\psi^\dag(x)\gamma_5\psi(x)$) profile, cf. Fig.~\ref{chiral}.
In the latter case, the chirality contained in the two peaks is of opposite sign.
These profiles provide a strong hint toward a picture where two zero modes with equal and
opposite chiralities interact weakly, becoming a pair of near-zero modes when properly superposed.

The fermion zero mode $\psi_0(x)$ associated with an instanton is known analytically~\cite{thooft2},
giving a density of the form
\begin{align}
    \label{eq:instantonsolution}
    \psi^\dagger_0(x) \psi_0(x)= \frac{2\rho^2}{\pi^2(x^2+\rho^2)^{3}}~,
\end{align}
where $\rho$ is the radius of the instanton. When three of the spacetime coordinates
are integrated over, the density along the remaining fourth coordinate, say $y$, becomes 
\begin{align}
    f_y(y) = \frac{\rho^2}{2(y^2+\rho^2)^{3/2}}~.
    \label{eq:instantonsolint}
\end{align}
An estimate of the instanton size $\rho$ can be obtained by either finding the distance
where this integrated density falls below $1/\sqrt{8}$ of its maximal value or by 
fitting Eq.~\eqref{eq:instantonsolint} to the measured density.
When estimating the size in the temporal direction,
the second approach has the advantage that it can also accommodate
cases where the size of the instanton $\sim 1/T$ and the 
periodic copies in the temporal direction have noticeable overlap.
This is achieved by replacing the fit function by $\sum_{k=-n}^n f_\tau(\tau+k/T)$.

At $1.5\,T_c$, we measured the wavefunction density of the zero modes along each 
coordinate direction for all configurations with $|Q|=1$ by summing over the other
three directions.
The radii of the profiles along the $x$, $y$ and $z$ directions, $\rho_x$, $\rho_y$ and
$\rho_z$, were averaged over to give a spatial radius of $\rho_\sigma=0.223(8)\,\mathrm{fm}$,
essentially independent of which method was used.
Along the temporal direction, using the adjusted fit ansatz we obtained a radius that was
only slightly larger, namely $\rho_\tau=0.24(1)\,\mathrm{fm}$.

\begin{figure}[t]
    \centering
    \includegraphics{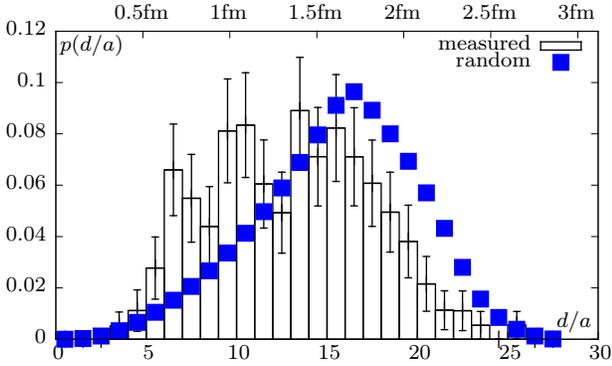}
    \caption{Distribution of distances between the instanton and antiinstanton that
    couple to give rise to a near-zero mode. The blue points show the expected
    distribution if the instanton and antiinstanton were to be distributed randomly and
    independently of each other on a $32^3\times 8$ lattice.}
    \label{fig:dist}
\end{figure}

In order to study the distribution of the distance between the
instanton and antiinstanton forming a pair, the lattice points with the lowest and highest 
chiral density of the corresponding near-zero modes were identified and the distance
between them was measured.
The result is shown in Fig.~\ref{fig:dist}. We compare this with the expected
distribution of the separation if an instanton and an antiinstanton were to be
distributed on the lattice randomly, independent of each other. It can be seen that
the measured distribution compares quite well with the random distribution
with a slight excess at small separations, further
supporting the picture that instantons and antiinstantons are weakly interacting.

\begin{figure}[t]
    \begin{center}
        \includegraphics[width=0.4\textwidth]{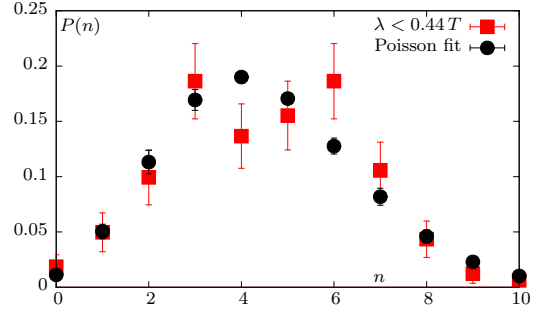}
        \caption{Configuration distribution of the total number $n$ of zero and
        near-zero modes at $1.5\,T_c$ and Poisson fit to the data.}
        \label{fig:smallmodesdist}
    \end{center}
\end{figure}

In a weakly interacting random ensemble of such topological objects, the 
assumption of independent occurrence results in a Poisson distribution for the total number $n$
of instantons and antiinstantons~,
\begin{align}
\label{eq:poisson}
P_\kappa(n)=e^{-\kappa}\kappa^n/n!\,,
\end{align}
where $\kappa$ is a parameter that is equal to the
ensemble average $\left<n\right>$ as well as the variance $\sigma^2$
and $n$ is obtained by counting the number of eigenvalues below a cutoff $\lambda_0$.
To fix the cutoff, we fitted Eq.~\eqref{eq:poisson} to the distribution of $n$
for different cutoffs and compared the $\chi^2$ per degree of freedom. The value
that was closest to 1, namely 1.03, was obtained for $\lambda_0=0.44\,T$.
The resulting distribution is shown in Fig.~\ref{fig:smallmodesdist} together
with the Poisson fit which gives $\kappa=4.5(2)$. At this cutoff, we obtained
$\left<n\right>=4.50(14)$ and $\sigma^2=4.2(4)$ from averaging over the configurations,
confirming that instantons and antiinstantons indeed occur almost independently. 

Taking 4.5(2) as the average total number of instantons and antiinstantons per 
configuration, we obtain a density of $0.147(7)\,\mathrm{fm}^{-4}$, which is much
lower than the value predicted from the strongly interacting dense ILM, $1\,\mathrm{fm}^{-4}$~\cite{shuryak}.
However, since we are at a temperature much higher than the chiral crossover temperature $T_c$,
it is not surprising that we get values expected rather from a dilute and weakly interacting
ensemble.

Altogether, the findings described in this section are giving support to a picture
in which for sufficiently high temperatures of $T\gtrsim1.5\,T_c$, the infrared behavior of
QCD can be described as that of a dilute gas of instantons and antiinstantons which are
interacting weakly with each other.

\subsection{The localization properties of the eigenmodes} \label{sc:localprop}

The profiles of the near-zero modes have suggested that these are localized structures. To
further quantify their localization properties and those of the bulk modes, we study the so-called
participation ratio (PR), defined for a normalized eigenvector $\psi(x)$ 
of the Dirac operator as
\begin{align}
    PR=\frac{1}{N_\sigma^3 N_\tau}\left[\sum_x \left(\psi^\dagger(x)\psi(x)\right)^2\right]^{-1}~.
\end{align}
It is the fraction of the total lattice volume occupied by the eigenmode. If the 
eigenvector is distributed equally on the entire four-volume, this quantity is unity.

\begin{figure}[t]
    \centering
        \includegraphics{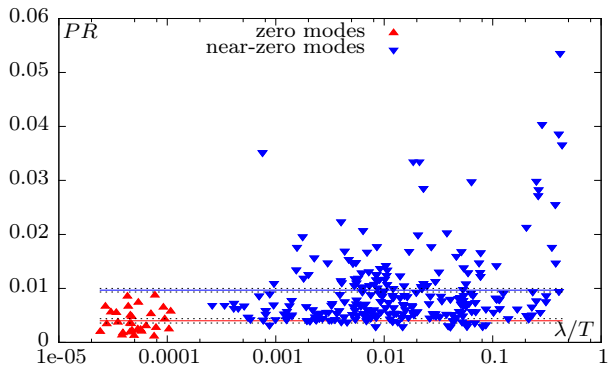}
    \caption{Participation ratio of zero modes of $|Q|=1$ configurations
        and near-zero modes ($\lambda/T<0.4$) as well as their
    average values as function of the eigenvalue $\lambda$ at $T\sim 1.5\,T_c$.}
    \label{fig:partratznz}
\end{figure}

First we use the PR  to corroborate our observation made in Sec.~\ref{sc:topobj}. If indeed
the near-zero modes represent weakly interacting instanton-antiinstanton
pairs, the PR of a typical near-zero mode should be about twice as large as that of
a zero mode. The comparison of the PR of the near-zero modes and of the zero modes
of $\vert Q\vert=1$ configurations is shown in Fig.~\ref{fig:partratznz}. The ratio
of the average PR of a near-zero and that of a zero mode indeed is $1.85$. The PR values of
near-zero modes fluctuate about the mean value, so not all of them support this
picture, but there is a significant fraction that does.

\begin{figure}[t]
    \begin{center}
        \includegraphics[width=0.4\textwidth]{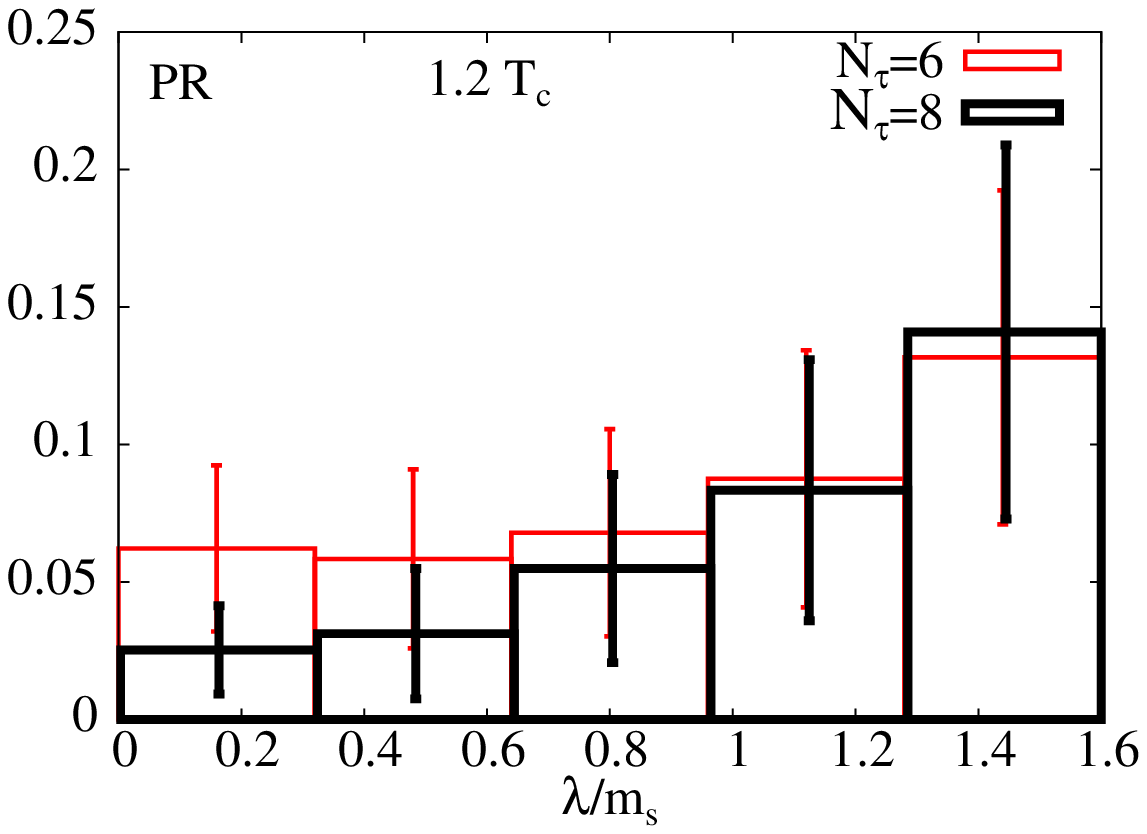}
        \includegraphics[width=0.4\textwidth]{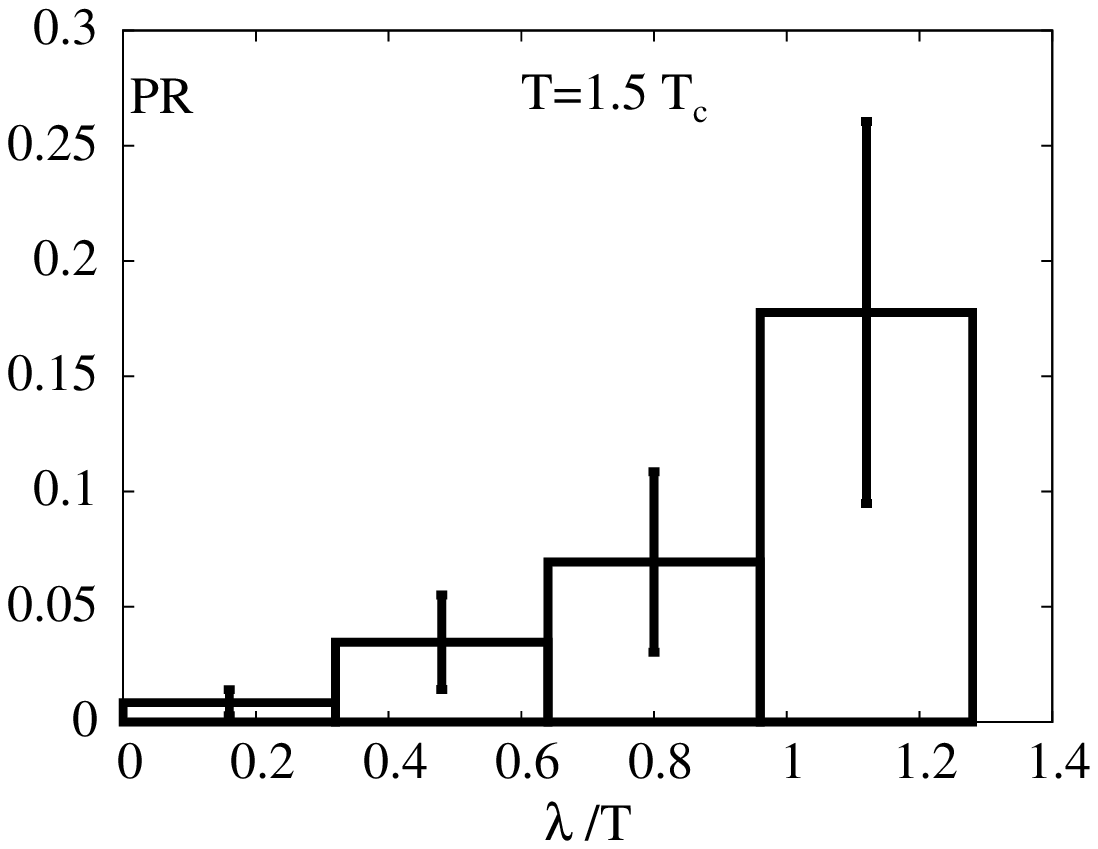}
        \caption{The PR for eigenvectors at 1.2 and 1.5 $T_c$. The 1.2 $T_c$ data are
        for two different lattice spacings to study the cutoff dependence of our
        results.}
        \label{fig:pr}
    \end{center}
\end{figure}

At $1.2\,T_c$, the PR histograms are measured for two values of the lattice
spacing and compiled in Fig.~\ref{fig:pr}. 
The low-lying eigenvalues are more localized than the bulk modes; however,
the PR gradually increases as one goes towards the bulk. At $1.5\,T_c$, the low-lying
modes below $\lambda\leq0.4\,T$ average to a PR value of about 0.01. As one enters the bulk eigenvalue
region for $\lambda>0.4\,T$, see for example Fig.~\ref{fig:eigval7},
there appears to be a rise of the PR. The value of
$\lambda\approx0.4\,T$ may thus be considered as a mobility edge separating 
the localized near-zero eigenstates from the delocalized bulk states. 
The presence of localized as
well as delocalized states is observed in disordered semiconductors whose dynamics
is described by the Anderson Hamiltonian. In the Anderson model, the electron states
at the band edge are localized whereas the states at the band center remain
delocalized within the lattice.  The corresponding eigenvalues of the Anderson
Hamiltonian change from Poisson statistics at the band edge to random matrix
theory statistics at the band center. 

A comparison with random matrix model predictions can be achieved by 
looking at the distribution of the distance $s$ between two consecutive eigenvalues,
i.e. the level spacing distribution.
In order to understand its universal properties, it is necessary to map the
eigenvalues onto new values using an unfolding procedure~\cite{guhr}, thereby
removing the non-universal global scale of the system. By construction, the 
unfolded eigenvalues that are obtained by this method have a mean level spacing of unity.

\begin{figure}
    \begin{center}
        \includegraphics[width=0.4\textwidth]{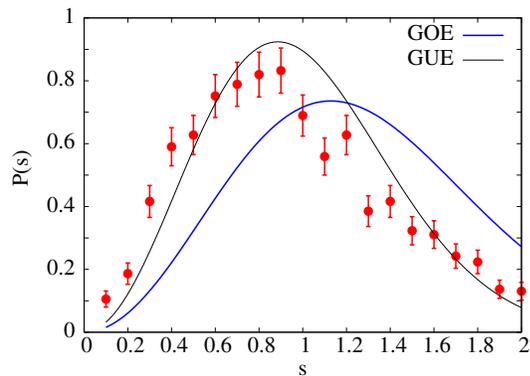}
        \caption{The universal level spacing distribution for $\lambda>0.4\,T$
            at $1.5\,T_c$ compared to the Random Matrix theory Gaussian 
        Unitary ensemble (GUE) and the Gaussian Orthogonal ensemble (GOE).}
        \label{unfold}
    \end{center}
\end{figure}

While the unfolded level spacings of the localized near-zero modes should 
follow a Poisson distribution because of their mostly independent occurrence,
the bulk modes should be strongly mixed. 
If the bulk is highly disordered, the corresponding level spacing distribution 
should follow the same distribution as the eigenvalues of a
random matrix of an appropriate symmetry group.  The Dirac operator for QCD  with 
matter fields in the fundamental representation falls into the same symmetry group as Gaussian 
Unitary ensembles (GUE) whereas  for the case of two colors it is in the same universality 
class as Gaussian Orthogonal ensembles (GOE).
 The level spacing distribution for the unfolded bulk eigenvalues 
of the overlap operator on HISQ configurations at $1.5\,T_c$ is shown in
Fig.~\ref{unfold}.  It shows an agreement with a random matrix theory with Gaussian unitary
matrices. This is in general agreement with similar studies of the localization of low-lying modes done before for the
quenched theory \cite{ehn,gg1,kovacs1,kovacs2} and also with dynamical staggered
fermions \cite{go,gg2,kovacs3} on smaller lattice sizes.

\section{Conclusions}    \label{sc:conclusion}
In this work we have investigated the temperature dependence of
the anomalous $U_A(1)$ symmetry breaking in the high temperature phase of QCD with 
two light quark flavors.
To this end we have employed the overlap Dirac operator exploiting its
property of preserving the index theorem even at nonvanishing lattice spacing.
We have applied the overlap operator on large volume HISQ gauge field configurations 
and computed its low-lying eigenmodes.
We observe the presence of zero as well as near-zero modes in the investigated
temperature range of $T_c\lesssim T \lesssim 1.5\,T_c$.
By comparing the low-lying eigenmodes from two lattice spacings and studying 
the effects of smearing we have shown that these infrared modes are not mere
lattice cutoff effects.

We mainly analyzed configurations which have been obtained at a light sea
quark mass corresponding to a pion mass of $160\,\mathrm{MeV}$. However,
within the set of configurations at our disposal, at a temperature near $T_c$
we could confirm the accumulation of the near zero eigenvalues also at
a quark mass considerably below its physical value.

By quantifying the contribution of the near-zero eigenmodes to a specific 
combination of two point correlation functions, $\chi_\pi-\chi_\delta$,  
we conclude that these modes are primarily responsible for the anomalous breaking
of the axial symmetry in QCD still being visible for $T_c\lesssim T\lesssim 1.5\,T_c$.
Through detailed studies of their spacetime profiles, localization properties and 
distributions over gauge configurations we have shown that for $T\sim1.5\,T_c$ 
the near-zero modes follow the behavior as expected of a gas of widely separated,
weakly interacting instantons and antiinstantons. At $1.5\,T_c$ we find the density of
(anti)instantons to be $0.147(7)\,\text{fm}^{-4}$, with a typical 
radius of $0.223(8)$ fm. At this temperature, the spatial volume of our lattice
was $\sim(3.3\,\text{fm})^3$ with $1/T\sim0.83\,\text{fm}$, suggesting that the
instanton gas is indeed dilute, the instanton size is smaller than $1/T$ and our chosen volume
being large enough to accommodate more that one instanton--antiinstanton pair on
average. In conclusion, our study suggests that at $T\sim1.5\,T_c$ the origin of
global $U_A(1)$ breaking in QCD is due to the dilute gas of weakly interacting 
instantons and antiinstantons. 

For an independent confirmation of our results, it would clearly be desirable to carry
out a similar analysis with dynamical chiral fermions. While the lattice spacing
effects in this work appear to be small, it will further be necessary to control
the subtle extrapolations to the continuum as well as the chiral limit in such a 
future investigation.

\begin{acknowledgments} 

This work has been supported in part through contract DE-SC0012704  with the U.S.
Department of Energy, the BMBF under grant 05P12PBCTA, EU under grants 238353 and 283286 
and the GSI BILAER grant. Numerical calculations
have been performed using GPU clusters at Bielefeld University. The GPU codes used 
in our work were in part based on some publicly available QUDA libraries~\cite{quda}.
SS would like to thank Gernot Akemann and Mario Kieburg for discussions and their helpful suggestions.
\end{acknowledgments} 

\end{document}